\documentclass[sigconf, nonacm]{acmart}
\AtBeginDocument{%
  \providecommand\BibTeX{{%
    Bib\TeX}}}

\newtheorem{example}{Example}
\usepackage{amsmath,amsfonts}

\usepackage{enumitem}
\usepackage{xcolor}
\usepackage{cleveref}
\usepackage{pifont}
\usepackage{caption}
\usepackage{graphicx}
\usepackage{balance}  
\usepackage[linesnumbered,vlined,ruled]{algorithm2e}
\usepackage{lipsum}
\usepackage{amsmath}
\usepackage{physics}
\usepackage{subfigure}
\usepackage{tabularx}
\usepackage{color}
\usepackage{colortbl}
\usepackage{float}
\usepackage{listings}
\usepackage{multirow}
\usepackage[normalem]{ulem}
 \usepackage{longtable}
\usepackage{enumitem}
\usepackage{multirow}
\usepackage{booktabs}
\usepackage{capt-of}
\usepackage{pifont}
\usepackage{ulem}
\usepackage{soul}
\usepackage{cancel}
\usepackage{bm}
\usepackage{url}
\usepackage{caption}
\usepackage{tikz}

\def\BibTeX{{\rm B\kern-.05em{\sc i\kern-.025em b}\kern-.08em
    T\kern-.1667em\lower.7ex\hbox{E}\kern-.125emX}}

 \graphicspath{{./Graph/}, {./Fig/}, {./Legend/}}

\long\def\comment#1{}

\newcounter{definition}[section]
\renewcommand{\thedefinition}{\nthesection.\arabic{definition}}
\newenvironment{definition}{
     \refstepcounter{definition}
     {\vspace{1ex} \noindent\bf  Definition  \thedefinition:}}{
     \vspace{1ex}} 

\newcounter{theorem}[section]
\renewcommand{\thetheorem}{\nthesection.\arabic{theorem}}

\newcounter{lemma}[section]
\renewcommand{\thelemma}{\nthesection.\arabic{lemma}}
        
\newcounter{proposition}[section]
\renewcommand{\theproposition}{\nthesection.\arabic{proposition}}

\newcounter{remark}[section]
\renewcommand{\theremark}{\nthesection.\arabic{remark}}

\newcommand{\nthesection}{\arabic{section}}

\newcommand{\stitle}[1]{\vspace{1ex} \noindent{\bf #1}}

\newcommand{\green}[1]{\textcolor{green}{}}

\newcommand{\kw}[1]{{\ensuremath {\mathsf{#1}}}\xspace}

\usepackage{algorithmic}
\usepackage{graphicx}
\usepackage{textcomp}
\usepackage{enumitem}
\usepackage{cleveref}
\usepackage{pifont}
\usepackage{cleveref}
\usepackage{makecell}
\usepackage{balance}

\crefformat{section}{\S#2#1#3} 
\crefformat{subsection}{\S#2#1#3}
\long\def\comment#1{}

\newcommand{\algocomment}[1]{\footnotesize $\rhd$ \emph{#1}}
\definecolor{LightCyan}{rgb}{0.88,1,1}
\definecolor{LightRed}{rgb}{1,0.88,1}
\definecolor{LightYellow}{rgb}{1,1,0.88}
\definecolor{LightGray}{gray}{0.8}

\newcommand{\token} {\tau}

\newcommand{\community} {\mathcal{C}}
\newcommand{\neighbor} {\mathcal{N}}

\newcommand{\Graph} {\mathcal{G}}
\newcommand{\VSet} {\mathcal{V}}
\newcommand{\ESet} {\mathcal{E}}
\newcommand{\ASet} {\mathcal{A}}
\newcommand{\SIM}{\text{sim}}
\newcommand{\Subgraph}{{\mathcal{S}}\xspace}
\newcommand{\OOM}{{OOM}\xspace}

\newcommand{\Fagg}{f_{\mathbb{A}}}
\newcommand{\Fcom}{f_{\mathbb{C}}}
\newcommand{\loss}{\mathcal{L}}

\newcommand{\Real} {\mathbb{R}}

\newcommand{\Reddit} {\kw{Reddit}}

\newcommand{\Cora} {\kw{Cora}}
\newcommand{\Citeseer} {\kw{Citeseer}}
\newcommand{\Cornell} {\kw{Cornell}}
\newcommand{\Texas} {\kw{Texas}}
\newcommand{\Washington} {\kw{Washington}}
\newcommand{\Wisconsin} {\kw{Wisconsin}}

\newcommand{\Product} {\kw{Amazon2M}}
\newcommand{\Orkut} {\kw{Orkut}}

\newcommand{\Fone} {\kw{F1}}
\newcommand{\Pre} {\kw{Pre}}
\newcommand{\Rec} {\kw{Rec}}

\newcommand{\AFC}{{{AFC}}\xspace}
\newcommand{\AFN}{{{AFN}}\xspace}
\newcommand{\EMA}{{{EQA}}\xspace}

\newcommand{\Transzero}{{{\sl TransZero}}\xspace}
\newcommand{\ATC}{{{\sl ATC}}\xspace}
\newcommand{\CTC}{{{\sl CTC}}\xspace}
\newcommand{\ACQ}{{{\sl ACQ}}\xspace}
\newcommand{\ALICE}{{{\sl ALICE}}\xspace}
\newcommand{\AQDGNN}{{{\sl AQD-GNN}}\xspace}
\newcommand{\QDGNN}{{{ \sl QD-GNN}}\xspace}
\newcommand{\ICSGNN}{{{\sl ICS-GNN}}\xspace}

\newcommand{\IACS}{{{\sl IACS}}\xspace}
\newcommand{\PLACE}{{{\sl PLACE}}\xspace}

\newcommand{\GIN}{{\sl GIN}\xspace}

\newcommand{\RGCN}{{\sl RGCN}\xspace}
\newcommand{\GAT}{{\sl GAT}\xspace}

\def\mathbi#1{\textbf{\em #1}}

\begin{document}

\title{PLACE: Prompt Learning for Attributed Community Search in Large Graphs}

\author{Shuheng Fang}
\affiliation{%
	\institution{Shenzhen Institute of Computing Sciences}
	\city{}
	\country{}%
}
\email{fangshuheng@gmail.com}

\author{Kangfei Zhao}
\affiliation{%
	\institution{Beijing Institute of Technology}
	\city{}
	\country{}
}
\email{zkf1105@gmail.com}

\author{Rener Zhang}
\affiliation{%
	\institution{Beijing Institute of Technology}
	\city{}
	\country{}
}
\email{larryzhang2004@outlook.com}

\author{Yu Rong}
\affiliation{%
	\institution{Chinese University of Hong Kong}
 \city{}
	\country{}
}
\email{yu.rong@hotmail.com}

\author{Jeffrey Xu Yu}
\affiliation{%
	\institution{The Hong Kong University of Science and Technology (Guangzhou)}
	\city{}
	\country{}
    }
\email{jeffreyxuyu@hkust-gz.edu.cn}

\begin{abstract}
Attributed Community Search (ACS) aims to identify communities in an attributed graph with structural cohesiveness and attribute homogeneity for given queries. 
While algorithmic approaches often suffer from structural inflexibility and attribute irrelevance, recent years have witnessed a broom in learning-based approaches that employ Graph Neural Networks (GNNs) to simultaneously model structural and attribute information. 
To improve GNN efficacy and efficiency, these approaches adopt various techniques to refine the input graph and queries, including query-dependent node pruning and feature enhancement. 
However, these refinements are either detached from the end-to-end optimization with the backbone model, or impose additional burdens on training resources, limiting their adaptability and practical usage in diverse graphs. 

In this paper, we propose \PLACE (\underline{P}rompt \underline{L}earning for \underline{A}ttributed \underline{C}ommunity S\underline {e}arch), an innovative graph prompt learning framework for ACS. 
Enlightened by prompt-tuning in Natural Language Processing (NLP), where learnable prompt tokens are inserted to contextualize NLP queries, 
\PLACE integrates structural and learnable prompt tokens into the graph as a query-dependent refinement mechanism, forming a prompt-augmented graph. 
Within this prompt-augmented graph structure, the learned prompt tokens serve as a bridge that strengthens connections between graph nodes for the query, enabling the GNN to more effectively identify patterns of structural cohesiveness and attribute similarity related to the specific query. 
We employ an alternating training paradigm to optimize both the prompt parameters and the GNN jointly. Moreover, we design a divide-and-conquer strategy to enhance scalability, supporting the model to handle million-scale graphs. Extensive experiments on 9 real-world graphs demonstrate the effectiveness of \PLACE for three types of ACS queries, where \PLACE achieves higher \Fone scores by {$14\%$} compared to the state-of-the-arts on average. 

\end{abstract}

\maketitle

\section{introduction}
\label{sec:introduction}

Attributed Community Search (ACS) is a fundamental graph analytic task aims at identifying communities in an attributed graph, which satisfy both structural cohesiveness and attribute homogeneity for a given query that consists of query nodes and query attributes. 
ACS plays a critical role in various real-world applications, including recommendation systems~\cite{yuan2019simple,tang2018personalized}, bioinformatics~\cite{lee2013hidden}, fraud detection~\cite{sarma2020bank} and social network analysis. ACS has been addressed using algorithmic approaches~\cite{ACQ,ATC, fang2020survey} following a two-stage paradigm. The first stage involves identifying candidate communities based on pre-defined structural patterns, such as $k$-core~\cite{li2015influential,sozio2010community,cui2014local}, $k$-truss~\cite{cs2,akbas2017truss} or $k$-clique~\cite{cui2013online,yuan2017index}. The second stage refines these candidate communities with attribute homogeneity constraints. 
However, these algorithmic approaches face significant limitations~\cite{AQDGNN,cgnp,ICSGNN}. They lack structural flexibility and attribute relevance, and fail to capture the joint correlations between structural and attribute information, which are essential for accurately identifying communities.

\begin{figure}[t!]
    \centering
    \includegraphics[width=0.4\textwidth]{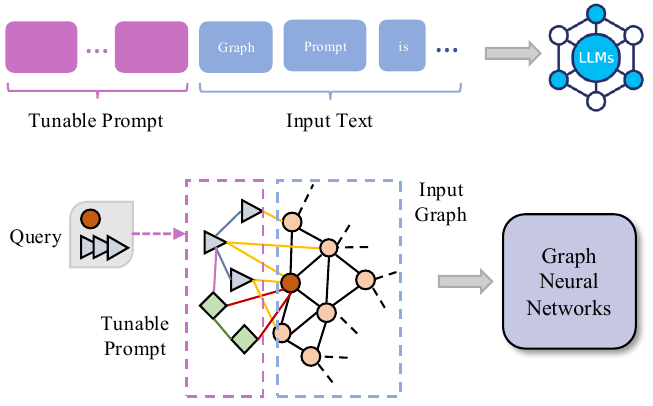}
    \hspace{-2ex}
    \caption{Query prompt inspired by the prompt of LLMs.}
    \label{fig:intro}
\end{figure}

To address the limitations of algorithmic approaches in the two-stage paradigm, learning-based approaches have been proposed~\cite{ICSGNN, ALICE, AQDGNN, IACS} that leverage Graph Neural Network (GNN) models~\cite{DBLP:journals/tnn/WuPCLZY21} to simultaneously capture structural and attribute information. 
These models are fed with the graph with its node features, the query encompassing query nodes and query attributes, and predict the likelihood of the community membership of each node w.r.t. the query.
To improve the efficacy and efficiency of the GNN, these approaches incorporate various techniques to refine the input graph and query,
which can be categorized into two types: decoupled refinement~\cite{ALICE, IACS, ICSGNN} and integrated refinement~\cite{AQDGNN}. 
Decoupled refinement preprocesses the graph or query using fixed node pruning or feature enhancement strategies. For instance, \ICSGNN~\cite{ICSGNN} and \ALICE~\cite{ALICE} adopt BFS and heuristic search to prune unpromising nodes in the graph, while \IACS~\cite{IACS} enhances the embeddings of queries through pre-trained graph embedding. 
However, these refinements are decoupled from the GNN model, missing the opportunity for end-to-end optimization, thus they may not be adaptable to any graph and query. 
The other category, integrated refinement introduces specialized neural network modules to enhance the query nodes and query attributes~\cite{AQDGNN},  which are collaboratively optimized with the backbone GNN. 
However, this refinement imposes extra burdens of computational resources on the ACS task and requires more training data to resist overfitting. 
Therefore, it is critical to introduce a flexible, adaptable, and lightweight refinement mechanism into ACS. 


Drawing the inspiration from prompt learning in the Natural Language Processing (NLP)~\cite{DBLP:conf/emnlp/LesterAC21} 
and Computer Vision (CV)~\cite{DBLP:conf/eccv/JiaTCCBHL22, DBLP:conf/cvpr/ParkB24} domains, we propose refining the input graph and ACS query through prompting techniques. 
In NLP tasks, prompt tokens are inserted beforehand the textual input, contextualizing it for foundation models as the illustration at the top of Fig~\ref{fig:intro}. The embeddings of these prompt tokens are fine-tuned to encapsulate task-level prior knowledge, facilitating the model to  achieve better reasoning for a specific task.
An envision arises that a graph prompt can  play a similar role as a prompt in NLP. 
The graph prompt will learn the latent knowledge of ACS query regarding both structural cohesiveness and attribute homogeneity, refining and guiding the input graph, query nodes and query attributes to generate more precise predictions. 
Compared with existing learning-based ACS approaches, the graph prompt for ACS serves as a more flexible refinement mechanism that injects the latent knowledge into the input graph and query without excessive computational overhead. 

Although prompting techniques have been introduced to the graph domain for typical graph learning tasks such as node-level and graph-level classification~\cite{DBLP:journals/corr/abs-2311-16534,DBLP:conf/kdd/SunZHWW22,gong2023prompt,shirkavand2023deep,zhu2023sgl,tan2023virtual}, distinctions of the task and objective incur new challenges in leveraging graph prompting for ACS. To be specific, unlike graph learning tasks where graph prompt is to unify the different learning tasks into the context of the pre-training tasks, graph prompt for ACS concentrates on enhancing the input graph and ACS query with learned prior knowledge, by unifying the input refinement step. As a query-specific binary classification task instead of classical node classification, query-specific prompt should be tailored, including how to design the prompt structure, how to insert the structural prompt into the input graph, and how to tune the prompt with the backbone ACS model. 
In contrast to the refinement of existing learning-based ACS, this query-specific prompt should enable end-to-end optimization and balance the computational overhead. 
The bottom part of Fig.~\ref{fig:intro} delineates a sketchy idea of this query-specific prompt for ACS. 

To fulfill this envision, we propose an innovative framework, named \underline{P}rompt \underline{L}earning for \underline{A}ttributed \underline{C}ommunity S\underline {e}arch (\PLACE). As illustrated at the bottom of Fig.~\ref{fig:intro}, by inserting learnable prompt into the graph, \PLACE enriches the input context, promoting the model to capture latent structural and attribute patterns regarding the query-dependent community. 
\PLACE integrates a complete solution of formulating the prompt and its structure, inserting the prompt, and tuning the prompt together with the model.
Precisely, the prompt tokens comprise attribute tokens and virtual node tokens, which are responsible for contextualizing the query attributes and query node respectively.
These prompt tokens are connected based on their similarity, forming a query-specific prompt graph that is seamlessly inserted into the input graph to fulfill input refinement and enhancement. 
\PLACE employs an alternating training paradigm to jointly optimize the prompt tokens and the ACS model, ensuring end-to-end optimization while balancing computational efficiency. 
In addition, to support large graphs efficiently, we design a divide-and-conquer strategy as a scale-up solution of \PLACE.

The contributions of this paper are summarized as follows:
\ding{182} We propose a new GNN-based ACS paradigm that first leverages query-specific prompts to handle ACS queries with multiple query nodes and multiple query attributes.
\ding{183} We devise a general framework named \PLACE that optimizes the query prompt and the GNN model jointly, and equip \PLACE with scalable training and inference algorithms to support ACS in million-scale graphs. 
\ding{184} We analyze the flexibility of \PLACE and its intrinsic connections and distinctions to existing GNN-based ACS approaches, revealing that the prompt mechanism of \PLACE simulates graph manipulations of structure and features.
\ding{185} We conduct comprehensive experiments on 9 real-world graphs for three types of ACS queries, demonstrating the effectiveness of \PLACE compared with 7 baselines.


\section{related work}
\label{sec:relatedwork}

\stitle{Attributed Community Search.}
Attributed community search (ACS) focuses on not only query nodes but also query attributes.
Various algorithmic approaches~\cite{ACQ, ATC,liu2020vac,guo2021multi,luo2020efficient} have been proposed that take both structural cohesiveness and attribute homogeneity into account.
For example, \cite{liu2020vac,guo2021multi,luo2020efficient} rely on their predefined subgraph patterns for community identification, which inherently limit their flexibility.
\ATC~\cite{ATC} and \ACQ~\cite{ACQ} are two representative approaches for ACS. 
Both of them adopt a two-stage process: first identifying potential candidate communities based on structural constraints, then computing attribute scores to verify the candidates. 
However, the independent two-stage process fails to capture the intrinsic correlations between structural  and attribute information in a joint manner, ultimately yielding suboptimal results. 

Database community has witnessed the recent boom in learning-based approaches for CS ~\cite{AQDGNN, ICSGNN,coclep,cgnp,communityAF,IACS,ALICE,wang2024efficient, DBLP:journals/tkde/SongZYWW24}.
However, among these approaches, only \AQDGNN, \ALICE and \IACS support multi-node queries and attributed queries. 
\AQDGNN~\cite{AQDGNN} proposes a GNN-based supervised model for ACS in a single graph. The model is trained by a collection of ACS queries with corresponding ground-truth, and predicts the communities for unseen queries.
The limitation of \AQDGNN lies in its lack of generalizability to new communities, graphs and attributes not encountered during training. 
\ALICE~\cite{ALICE} employs a two-phase strategy; first extracting promising candidate subgraphs and then identifying communities by a consistency-aware network. 
\IACS~\cite{IACS} introduces a `training-adaptation-inference' 
workflow that enables swift model adaptation to new graphs and communities, enhancing flexibility and applicability.

\stitle{Graph Prompt Learning.}
The general idea of prompt learning is to reformulate downstream tasks into the context of pre-training tasks, effectively bridging the gap between the downstream and pre-training tasks.
Drawing inspiration from prompting techniques in NLP and CV, graph prompt learning has become the cornerstone of the `pre-training and prompting' paradigm in the realm of graph learning, as comprehensively reviewed in a recent survey~\cite{DBLP:journals/corr/abs-2311-16534}.
As an early study, GPPT~\cite{DBLP:conf/kdd/SunZHWW22} incorporates learnable graph label prompts that transform the node classification task into a link prediction task.
GraphPrompt~\cite{DBLP:conf/www/LiuY0023} unifies node-level and graph-level tasks into a subgraph similarity learning task and employs a learnable prompt to assist the downstream tasks.
All-in-One~\cite{DBLP:conf/kdd/SunCLLG23} introduces a generic graph prompt learning framework, including prompt token, token structure, and insert pattern, unifying the node-level, edge-level, and graph-level tasks under a single paradigm. 
GCOPE~\cite{DBLP:conf/kdd/ZhaoCS0L24} leverages virtual node prompt tokens to align diverse structural patterns and semantic features across different graph datasets, enhancing cross-domain adaptability.

\section{preliminaries}
\label{sec:preliminaries}
We first provide the problem statement of ACS (\cref{sec:pre:definition}) and then present the background of learning-based ACS by GNN (\cref{sec:pre:gnn}). Finally, we introduce graph prompt as our technical foundation (\cref{sec:pre:prompt}).

\comment{
\begin{table}[t]
\footnotesize
\centering
\caption{Frequently-used Notations}
\label{tab:notation}
\begin{tabular}{c|c}
\toprule
Notation & Description                 \\\midrule
$\Graph = (\VSet, \ESet, \ASet)$ & an attributed graph \\
$ q = (\VSet_q, \ASet_q)$        & a query with node set $\VSet_q$ and attribute set $\ASet_q$\\
$\community_q=(\VSet_{\community}, \ESet_{\community}, \ASet_{\community})$ & the community containing query $q$\\
$n/m/a$  &number of nodes/edges/attributes of graph $\Graph$\\
$h^{(k)}(v)$  & the embedding of node $v$ in $k$-th layer\\
$\VSet_{s{(l)}}$ &the candidate node set sampling in $l$-th layer\\
$\token_q$ & prompt tokens for query $q$\\
$\Graph_p=(\VSet_p,\ESet_p)$ & prompt graph \\
$\Graph_m=(\VSet_m,\ESet_m)$ & prompt-augmented graph\\

 \bottomrule
\end{tabular}
\end{table}
}

\begin{figure*}[t]
    \centering
    \includegraphics[width=0.8\linewidth]{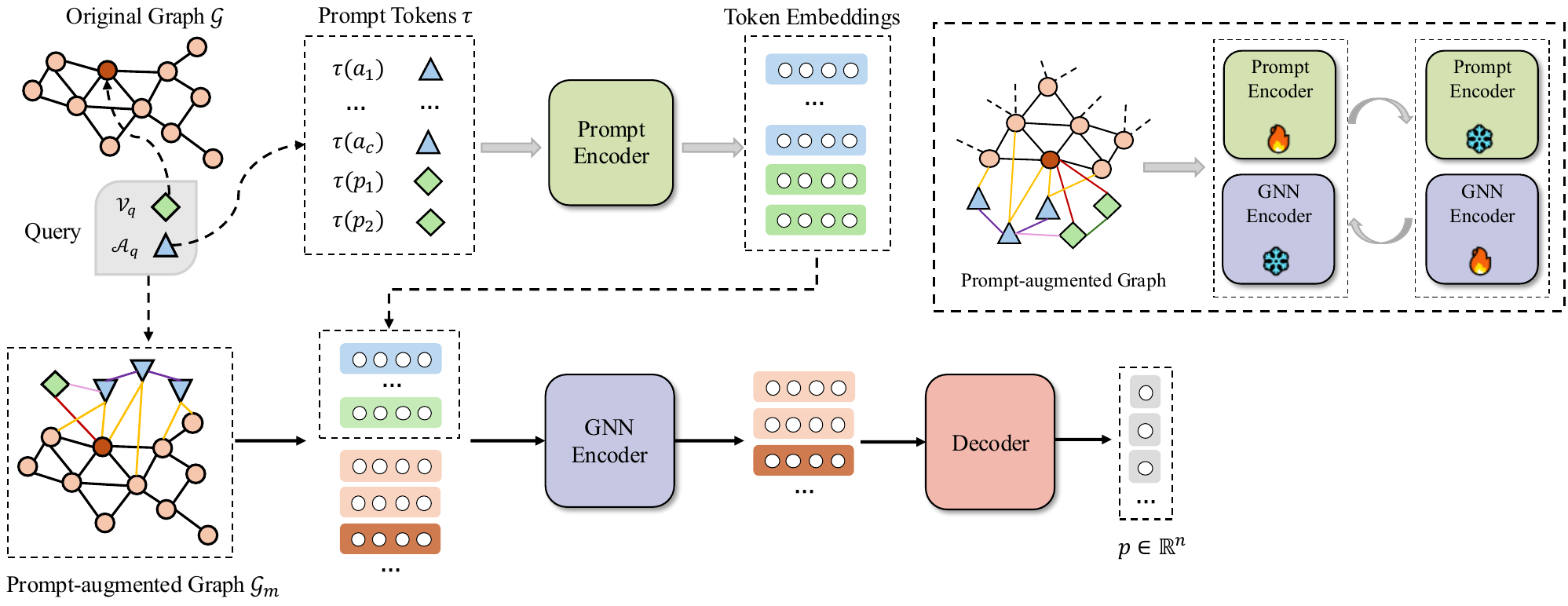}
     \caption{Architecture and Training Process of \PLACE}
    \label{fig:architecture}
\end{figure*}

\subsection{Problem Statement}
\label{sec:pre:definition}
An undirected attributed graph $\Graph = (\VSet, \ESet, \ASet)$ consists of a set of nodes $\VSet$, a set of undirected edges $\ESet \subseteq \VSet \times \VSet$, and a set of node attributes $\ASet$. 
Let $n = |\VSet|$, $m = |\ESet|$ and $c =|\ASet|$ be the number of nodes, edges, and attributes, respectively. 
Each node $v_i$ possesses its attribute set $\ASet_i$, and $\ASet$ is the union of all the node attribute sets, i.e.,
$\ASet = \ASet_1 \cup \cdots \cup \ASet_n $. 
The neighborhood of node $v_i$ is denoted as $\mathcal{N}(v_i)=\{v_j \mid(v_j,v_i)\in \ESet\}$.
A community in $\Graph$ is a cohesive subgraph, denoted as $\community=(\VSet_{\community}, \ESet_{\community})$, where $\VSet_{\community} \subseteq \VSet$ and $\ESet_{\community} = \{(v_i, v_j) \mid v_i, v_j \in \VSet_{\community}, (v_i, v_j) \in \ESet\}$. 
 
\stitle{Attributed Community Search (ACS).} For an attributed graph $\Graph = (\VSet, \ESet, \ASet)$, given a query $ q = (\VSet_q, \ASet_q)$ where $\VSet_q \subseteq \VSet$ is a set of query nodes and $\ASet_q \subseteq \ASet$ is a set of query attributes, the problem of {Attributed Community Search}~(ACS) aims to find the query-dependent community $\community_q=(\VSet_{\community_q}, \ESet_{\community_q})$. Nodes in community $\community_q$ need to be structurally cohesive and attribute homogeneous simultaneously, i.e., the nodes in the community are densely intra-connected as $|\ESet_{\community_q}| \gg |
\{(v_i, v_j)\mid v_i \in \VSet_{\community_q}, v_j \in \VSet \setminus \VSet_{\community_q}\}|$, 
and the attributes of these nodes are similar. 
The community $\community_q$ in this paper is not restricted in any $k$-related
subgraph, instead it is learned from given community membership
ground-truth.

\stitle{Learning-based ACS.} The general process of the learning-based approaches~\cite{ICSGNN,AQDGNN,coclep,cgnp,communityAF} consists of two stages: the training stage and the inference stage. In the training stage, for a graph $\Graph$ with $n$ nodes, a parametric ML model $\mathcal{M}: q \mapsto [0, 1]^n$ is constructed offline from a set of queries and corresponding ground-truth communities.
In the inference stage, for an online new query,  the model $\mathcal{M}$ predicts the likelihood of whether each node is in the community of the query, as a vector $\hat{y} \in [0, 1]^n$.
Distinguished from prior learning-based approaches, in this paper, we consider utilizing query-specific prompts to enrich the prediction context of the model. For a graph $\Graph$, model $\mathcal{M}: (q, \token_q) \mapsto [0, 1]^n$ is trained offline over a set of queries and their ground-truth communities, and then deployed for online prediction. Here, $\token_q$ is a learnable prompt of query $q$, analogous to the query prompt of foundation models for NLP tasks.

The query supported can be non-attributed queries ($q = (\VSet_q, \emptyset)$) for CS and attributed queries ($ q = (\VSet_q, \ASet_q)$) for ACS.
To be concise, we consider the ACS problem in this paper and regard CS as a special case of ACS ($\ASet_q = \emptyset$).
Distinguished from algorithmic approaches, the community $\community_q$ discovered by learning-based approaches is not restricted to any specific $k$-related subgraph. 


\subsection{GNN for Learning-based ACS}
\label{sec:pre:gnn}
Existing learning-based approaches~\cite{ICSGNN,AQDGNN,coclep,cgnp,communityAF,IACS} employ GNN as their backbone models. A GNN of $K$-layer follows a neighborhood aggregation paradigm to generate a new embedding for each node by aggregating the embeddings of its neighbors in $K$ iterations.
Specifically, let $h^{(k)}(v)$ denote the $d$-dimensional embedding of node $v$ in the $(k)$-th iteration (layer). 
In the $k$-th layer, an aggregate function $\Fagg^{(k)}$ aggregates the embeddings of the neighbors of $v$ output in $(k - 1)$-th layer as Eq.~\eqref{eq:gnn:fagg}. Subsequently, a combine function $\Fcom^{(k)}$ updates the embedding of $v$ by combining the aggregated embeddings in $k$-th layer and the embedding of $v$ in $(k-1)$-th layer as Eq.~\eqref{eq:gnn:fcom}.   
\begin{align}
	a^{(k)}(v) &= \Fagg^{(k)}(\{ h^{(k - 1)}(u) \mid u \in \neighbor(v)\}), \label{eq:gnn:fagg} \\
	h^{(k)}(v)  &= \Fcom^{(k)}(h^{(k - 1)}(v), a^{(k)}(v)). \label{eq:gnn:fcom}
\end{align}
The aggregate and combine functions of each layer, $\Fagg, \Fcom: \Real^d \rightarrow \Real^d$, are neural networks. 
The initial node embedding, $h^{(0)}(v)$, encapsulates the features of node $v$ and the identifier of whether $v$ is in the query node set. 
Through the transformation of $K$ layers, GNN-based models predict the likelihood that node $v$ belongs to the community of the query using a prediction layer $\hat{f}: \Real^{d} \rightarrow \Real$ as shown in Eq.~\eqref{eq:gnn:predict}, followed by sigmoid activation $\sigma$. 
\begin{align}
    \label{eq:gnn:predict}
     \hat{y}(v) = \sigma\left(\hat{f}(h^{(K)}(v))\right)
\end{align}
Given a set of training queries $Q$ and ground-truth $L$, ACS is formulated as a query-specific binary classification task in $\Graph$, where GNN-based models are trained by minimizing the binary cross entropy (BCE) loss, as shown in Eq.~\eqref{eq:loss:bce}.
\begin{align}
	\label{eq:loss:bce}
	\loss(q; \theta) = - \sum_{v^{+} \in l_q^+} \log\hat{y}(v^+) -  \sum_{v^{-} \in l_q^-} \log (1-\hat{y}(v^-)) 
\end{align}
%
In the test stage, for an online query, the model $\mathcal{M}$ first predicts the probability of whether each node is in the community of the query, as a vector $\hat{y} \in [0, 1]^n$. Then, a threshold on $\hat{y}$ is used to determine the membership of the community.


\subsection{Graph Prompt}
\label{sec:pre:prompt}
The graph prompt mechanism is composed of three elementary components: (1) prompt token, (2) token structure, and (3) insertion patterns~\cite{DBLP:conf/kdd/SunCLLG23}.
Given a graph $\Graph=(\VSet,\ESet)$, prompt tokens are a set of vectors $\bm{\token} = \{ \token_1, \cdots, \token_{|\bm{\token}|}\}$, where each token $\token_i$ is a learnable embedding of $d$-dim, the same size as the node features of $\Graph$. 
The token structure is the pairwise relationship among tokens, denoted as $\{ (\token_i, \token_j)\mid \token_i, \token_j \in \bm{\token}\}$, which differs from the linear token structure in the NLP domain. 
An edge $(\token_i, \token_j)$ exists iff $\SIM(\token_i,\token_j) > \delta$, where $\SIM(\cdot, \cdot)$ is a similarity function, i.e., inner product followed by the sigmoid function. 
Thereby, a prompt graph $\Graph_p=(\VSet_p,\ESet_p)$ is constructed, where $\VSet_p$ is the set of prompt tokens and $\ESet_p$ is the set of edges among tokens. 
For the insertion pattern,  an inserting function $\Psi$ is defined that inserts the prompt graph $\Graph_p$ into the original graph $\Graph$, forming a prompt-augmented graph as $\Graph_m=\Psi(\Graph,\Graph_p)$. The prompt-augmented graph is finally processed by GNN-based models for training and prediction. 
In this paper, we extend the graph prompt mechanism for dealing with ACS, where each ACS query is combined with a structural prompt as the input context of the model, enabling the model to better capture structural cohesiveness and the attribute similarity of the query.


\section{The architecture of \PLACE}
\label{sec:method}
In this section, we present the overall architecture of \PLACE, which comprises the query-specific graph prompt and the GNN-based model for ACS.
Fig.~\ref{fig:architecture} illustrates the general process of how \PLACE handles an ACS query $q=(\VSet_q,\ASet_q)$ in a graph $\Graph$ and the alternative training process. Regarding the query $q$, a query-prompt graph is constructed and inserted into $\Graph$, forming a prompt-augmented graph $\Graph_m$. Subsequently, this prompt-augmented graph $\Graph_m$ is then fed into an encoder-decoder GNN-based model to predict the community membership for each node in $\Graph$. 
The core design of \PLACE lies in the construction of the query-prompt graph as well as the prompt-augmented graph, which encapsulate the enhanced, contextualized, and customized information of query $q$ for a graph $\Graph$. Notably, the node features and the topological structure of the query-prompt graph are learnable, which is jointly optimized together with the GNN-based model. 
This enables \PLACE to achieve powerful flexibility and generalization capability. 
We will elaborate on the construction and insertion of the query-prompt graph (\cref{sec:method:prompt}) and the GNN-based model (\cref{sec:method:model}) in the remainder of this section. The training algorithm of \PLACE and its scale-up solution are deferred to \cref{sec:training}.

\subsection{Query-Prompt Graph Construction}
\label{sec:method:prompt}
We introduce the query-prompt graph for an ACS query, analogous to a prompt for a natural language question, which can be inserted into the original graph. 
This enables the original graph to assimilate relevant information about the target community in response to the query, facilitating an accurate prediction for ACS.

\stitle{Query Prompt Tokens.}
For a graph $\Graph$, we initialize a set of prompt tokens $\bm{\token}=\{\token(a_1),\token(a_2)\dots, \token(a_c), \token(p_1), \token(p_2) \dots, \token(p_{v_n})\}$, where $\{ \token(a_i) \mid i=1 \cdots,c \}$ are $c$ tokens correspond to $c$ query attributes ${a_i \in \ASet}$ respectively, and $\{ \token(p_j) \mid j=1 \cdots, v_n\}$ are $v_n$ tokens correspond to $v_n$ virtual nodes.
Specifically, $\token(a_i), \token(p_j)$ are $d$-dimensional learnable vectors, which align to the dimension of node features. 
The number of virtual node tokens, $v_n$, is a hyper-parameter.
Each query $q=(\VSet_q,\ASet_q)$ identifies its query prompt tokens $\token_q$, formulated as Eq.~\eqref{eq:query_prompt}:
\begin{align}
    \label{eq:query_prompt}
    \bm{\token}_q = \{ \token(a_i) \mid a_i \in \ASet_q\} \cup \{ \token(p_j) \mid j=1 \cdots, v_n\},
\end{align}
which contains the prompt tokens of the query attributes as well as all the virtual  node tokens.
The virtual node tokens will encode structural patterns related to the query node set $\VSet_q$, while the query attribute tokens refine and enrich the features of the query attribute set $\ASet_q$, with each token representing a distinct query attribute in $\ASet_q$, by connecting to a query-prompt graph and inserting into the original graph as follows. 

\begin{definition}(Query-prompt Graph) Given a query $q=(\VSet_q,\ASet_q)$, the query-prompt graph of $q$ is an undirected graph $\Graph_p=(\VSet_p,\ESet_p)$ with node set $\VSet_p =\bm{\token}_q$ and edge set
\begin{align}    
\label{eq:query_prompt_graph}
\ESet_p &= \{ (\token(i), \token(j)) \mid \token(i), \token(j) \in \VSet_p \wedge \SIM(\token(i), \token(j)) > \delta \}. 
\end{align}


\end{definition}
With a little abuse of notation, we use the query prompt tokens to represent the nodes in $\VSet_p$ whose features are the embeddings of the tokens.
Given the two types of nodes in $\VSet_p$, there are three types of edges in $\ESet_p$, the edges between query attribute tokens, edges between virtual node tokens, and edges between two types of nodes. As shown in Eq.~\eqref{eq:query_prompt_graph}, an edge will connect two nodes if and only if the similarity of the two nodes surpasses a threshold $\delta$, where $\SIM(\cdot)$ is the similarity function. 

\stitle{Insertion to the Original Graph.}
Analogous to inserting a textual prompt into an NLP question, the query-prompt graph will be inserted into the original graph $\Graph$ to enhance the prediction context. 
We formulate the resulting graph as a prompt-augmented graph:

\begin{definition}(Prompt-augmented Graph): Given a query $q=(\VSet_q,\ASet_q)$ on graph $\Graph=(\VSet, \ESet, \ASet)$, the query-augmented graph of $q$ is an undirected graph $\Graph_m=(\VSet_m,\ESet_m)$ with node set $\VSet_m = \VSet_p \cup \VSet$ and edge set
\begin{align}  
\label{eq:query_prompt_graph_augmented}
\ESet_m &= \ESet_p \cup \ESet 
\cup \{ (\token(a_i), v) \mid a_i \in \ASet(v) \wedge v \in \VSet \} \\
&\cup \{ (\token(p_j), v) \mid p_i=1,\cdots, v_n, \wedge v \in \VSet_q \}. \nonumber
\end{align}
\end{definition}
Here, the node set $\VSet_m$ is the union of $\VSet$ and the query-involved prompt token set $\VSet_p$. 
Apart from the two edge sets of the original graph and the query-prompt graph, the edge set $\ESet_m$ also contains edges between the prompt tokens and original nodes.  
To be specific, as shown in Eq.~\eqref{eq:query_prompt_graph_augmented}, for query attribute tokens $\token(a_i)$, a node in $\VSet$ is connected to $\token(a_i)$ iff the node possesses the attribute $a_i$. 
For virtual node tokens, we establish full connections between them to each query node in $\VSet_q$.
We introduce the insertion function $\Psi$ to define how the query-prompt graph $\Graph_p$ is inserted into the original graph $\Graph$, such that $\Graph_m=\Psi(\Graph,\Graph_p)$. 
An example demonstrates the construction and insertion of a query-prompt graph.

\begin{figure}[t]
    \centering
\includegraphics[width=0.9\linewidth]{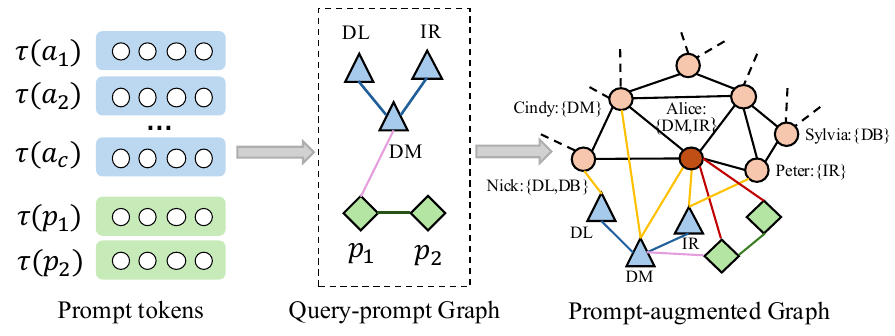}
    \caption{Construction  query-prompt graph given a query $q = (\VSet_q =\{\text{`Alice}'\},\ASet_q = \{\text{`DL'},\text{`DM'}, \text{`IR'}\})$}
\label{fig:exm:qpgraph}
\end{figure}

\begin{example}
    Given an original graph $\Graph$ and a query $q=(\VSet_q,\ASet_q)$, we construct a query-prompt graph as illustrated in Fig.~\ref{fig:exm:qpgraph}. 
    Suppose the prompt tokens $\bm{\tau}$ contain $c$ attribute tokens and 2 virtual node tokens. 
    The query $(\VSet_q =\{\text{`Alice'}\},\ASet_q = \{\text{`DL'},\text{`DM'}, \text{`IR'}\})$ identifies its prompt tokens as: $\token_q=\{\token(\text{`DL'}),\token(\text{`DM'}),\token(\text{`IR'}), \allowbreak \token(p_1),\token(p_2)\}$, as the node set $\VSet_p$ in the query-prompt graph.
    The edge set $\ESet_q$ is determined by the pairwise similarity, i.e., the inner product followed by a sigmoid function, across these tokens. For example, the embedding of $\token(\text{`DL'})$ and $\token(\text{`DM'})$ are $[1,0,0]$ and $[0.5,0,0.5]$, respectively, and the similarity threshold $\delta$ is $0.6$, we have $\SIM(\token(\text{`DL'}),\token(\text{`DM'}))=\sigma\left( \langle \token(\text{`DL'}),\token(\text{`DM'}) \rangle \right)  = \sigma(0.5)\approx  0.62>\delta$,
    and there is an edge $(\token(\text{`DL'}),\token(\text{`DM'})) \in \ESet_p$.
    Then, we insert query-prompt graph into the original graph $\Graph$ by connecting the tokens to their  corresponding nodes in $\Graph$.
 Specifically, token $\token(\text{`DM'})$ connects to `Alice' due to its association to `DM', and the  remaining attribute tokens in $\token_q$ connect to their corresponding nodes similarly. Virtual node tokens $\token(p_1)$, $\token(p_2)$ also connect to the query node `Alice'. This process yields the prompt-augmented graph, which combines the structural and attribute information as query context.
\end{example}

\subsection{GNN-based Model}
\label{sec:method:model}
After constructing the prompt-augmented graph $\Graph_m$ for a query $q$, we feed $\Graph_m$ into a GNN-based model for predicting the community membership of $q$. As illustrated in Fig~\ref{fig:architecture}, \PLACE adopts an encoder-decoder neural network model to process $\Graph_m$, which is formulated as Eq.~\eqref{eq:encoder-decoder} 
\begin{align}
    \label{eq:encoder-decoder}
    p(y_q \mid \Graph_m) = \rho\left(q, \phi_\theta(\Graph_m)\right),
\end{align}
where $\phi_\theta$ is a GNN-based encoder that generates a node embedding of $\Graph_{m}$, and $\rho$ is a parameter-free inner product decoder.  
Recall that in $\Graph_m$, the information of query $q$ is enriched by the insertion of the query-prompt graph. The initial node embedding of $\Graph_m$ contains the initial node embeddings in the original graph, denoted as $H^{(0)} = \{h^{(0)}(v) \mid v \in \VSet\} \in \Real^{n \times d}$, and the query-related node tokens $\tau_q \in \Real^{|\VSet_p|\times d}$. We use the original node features as the initial node embeddings of $H^{(0)}$, and all the attribute tokens $\tau(a)$ and virtual node tokens $\tau(p)$ are $d$-dim learnable vectors. 
A $K$-layer GNN, serving as the encoder $\phi_{\theta}$, fuses the embeddings of nodes and prompt tokens by message propagation as Eq.~\eqref{eq:gnn:fagg}-\eqref{eq:gnn:fcom}. The transformation is similar to the prompt-tuning process~\cite{DBLP:conf/acl/LiuJFTDY022} in foundation models, whose building blocks, the self-attention module~\cite{DBLP:conf/nips/VaswaniSPUJGKP17}, are a special type of GNN.
We extract the transformed embedding of the nodes in the original graph, denoted as $H^{(K)} \in \Real^{n \times d}$, as the output of the encoder. 
        

The decoder is an inner product operation that computes the probabilistic community membership of each node regarding query $q$, based on the similarity of the node embeddings $H^{(K)}$ and the embedding of query node $h_q$. The computation is formally represented as Eq.~\eqref{eq:decoder:predict}
\begin{align}
    \label{eq:decoder:predict}
	p(y_q \mid \Graph_m) = \sigma \left( \langle h_q, H^{(K)} \rangle \right) 
\end{align}
Here, we compute the query node embedding $h_q$ by the average of all the individual query node embeddings, i.e., $\frac{1}{|\VSet_q|}\sum_{v\in \VSet_q} h^{(K)}(v)$. $\langle,\rangle$ and $\sigma$ are the inner product operation and sigmoid function, respectively. 
The inner product operation indicates that the smaller the angle of $h_q$ and $h^{(K)}(v)$, the more likely $v$ and $\VSet_q$ belong to the same community.
Finally,  a threshold of 0.5 is applied on $p(y_q \mid \Graph_m) \in [0, 1]^n$ to extract the nodes in $\community_q$. 


\section{Training and Inference of \PLACE}
\label{sec:training}
In this section, we delve into the training algorithm of \PLACE (\cref{sec:training:plain}), and propose a strategy to scale up \PLACE to large graphs (\cref{sec:training:scale}). 
\subsection{Training of GNN and Prompt}
\label{sec:training:plain}
The training of \PLACE aims to optimize the parameters of both the GNN-based model $\theta$ and the prompt tokens $\bm{\token}$. Specifically, the model should be trained to perform the ACS tasks precisely, and the prompt should be tuned to navigate and supply relevant information for ACS queries. Thereby, 
given a set of training queries $Q=\{q_1, \cdots, \}$ with corresponding ground-truth $L = \{l_q, \cdots, \}$, the learning objective of \PLACE is formulated as Eq.~\eqref{eq:place:objective}, where the loss of a query is the negative log-likelihood of the predicted community membership as Eq.~\eqref{eq:place:loss}, equivalent to the BCE loss of Eq.~\eqref{eq:loss:bce}.  
\begin{align}
     \theta, \bm{\tau} &= \arg\min_{\theta, \bm{\tau}} \sum_{q \in Q}\loss(q, \tau_q; \theta) \label{eq:place:objective} \\
    \loss(q, \tau_q ; \theta)  &= -\log p(y_q \mid \Graph_m)  \label{eq:place:loss} 
\end{align}
To achieve this optimization objective, we use stochastic gradient descent to optimize $\bm{\tau}$ and $\theta$ alternatively, whose details are shown in Algorithm~\ref{alg:train}.
First, the parameters of the GNN and prompt tokens are initialized in line~\ref{line:train:initial}. 
In line~\ref{line:train:epoch:start}-\ref{line:train:epoch:end}, the algorithm iterates on each query to update the parameters of prompt tokens (line~\ref{line:train:prompt:start}-\ref{line:train:prompt:end}) or GNN model (line~\ref{line:train:gnn:start}-\ref{line:train:epoch:end}) alternatively. 
When training prompt tokens, we freeze the parameters of the GNN model, $\theta$, in line~\ref{line:train:prompt:start}, and conduct a forward pass of the model by Algorithm~\ref{alg:foward}.
With the returning predictive probability $p$, we compute the loss in line~\ref{line:train:prompt:loss}, and the parameters of prompt tokens w.r.t. query $q$ are updated by one step using the gradient of $\token_q$ in line~\ref{line:train:prompt:end}.
Afterwards, to train the parameters of the model, the parameters of the prompt tokens are frozen while those of the model are unfrozen in line~\ref{line:train:gnn:start}. Then, a similar forward propagation and loss computation are performed (line~\ref{line:train:gnn:start}-\ref{line:train:gnn:loss}), but the difference is that the parameters of the model are updated using the gradient of $\theta$ (line~\ref{line:train:epoch:end}). 
It is worth noting that the structure of a query-prompt graph w.r.t. a query may change across different epochs during training, as the parameters of $\bm{\token}$ change. The structure tends to converge at the end of the training.

\comment{
\begin{figure}[t]
    \centering
   \includegraphics[width=0.9\linewidth]{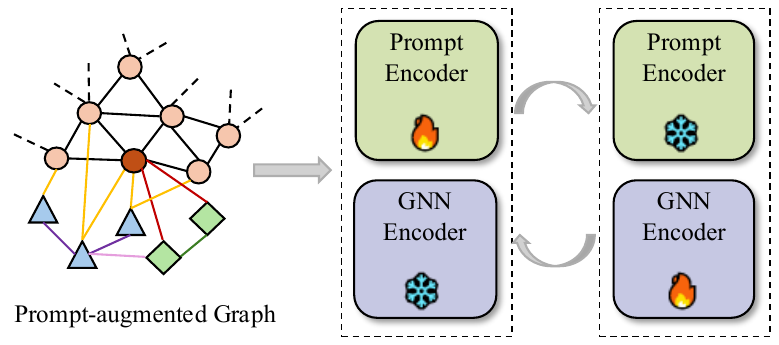}
    \caption{Alternative Training of Prompt Encoder and GNN}
    \label{fig:workflow}
\end{figure}}

\comment{
\begin{algorithm}[t]
    \caption{\PLACE Alternating Training Process}
	\label{alg:training}
	\DontPrintSemicolon
	\SetKwData{Up}{up}  \SetKwInOut{Input}{Input} \SetKwInOut{Output}{Output}
	\Input{input graph $\Graph$, training queries $Q = \{ q_1, \cdots\}$, number of epochs $T$, learning rates $\alpha, \beta$}
	\Output{parameters of ACS model $\theta$ and prompt tokens $\tau$}
	\SetKwFunction{Emit}{Emit}
	\SetKwFunction{Check}{Check}
        Initialize parameters $\theta$ and $\token$;\label{line:encoder:initial} \;
        \For{$epoch \leftarrow 1$ to $T$}{\label{line:promptepoch:start}
        Unfreeze $\tau$ and freeze $\theta$;  \label{line:prompt:freeze} \;
		\For { $q \in Q$ } 
		{ \label{line:prompt:start}  
            Construct the prompt graph $\Graph_p$ by extracting and connecting query-prompt tokens $\token_q$ of $q$;\label{line:prompt:constructgp}\;
		$\Graph_m\leftarrow\Psi(\Graph,\Graph_p)$;  \label{line:prompt:constructgm}  
            \algocomment{construct prompt-augmented graph} \;
		$p \leftarrow \rho(q, \phi_\theta(\Graph_m))$; \label{line:prompt:predict} 
            \algocomment{model forward pass}  \;
		Compute the loss $\mathcal{L}(q, \tau_q;\theta)$ by Eq.~\eqref{eq:place:loss}; \label{line:prompt:loss} \;
			
		$\token_q \leftarrow \token_q-\beta \nabla_{\token_q} \mathcal{L}(q)$; \label{line:prompt:end}  
    \algocomment{update parameters of prompt tokens}\; 
	}
    Freeze $\token$ and unfreeze $\theta$ ;  \label{line:GNN:unfreeze} \;
    \For { $q \in Q$ } 
		{ \label{line:GNN:start} 
            Construct the prompt graph $\Graph_p$ by extracting and connecting query-prompt tokens $\token_q$ of $q$; \label{line:GNN:constructgp}\;
		$\Graph_m \leftarrow \Psi(\Graph,\Graph_p)$;  \label{line:GNN:constructgm}  
            \algocomment{construct prompt-augmented graph} \;
		$p \leftarrow \rho(q, \phi_\theta(\Graph_m))$; \label{line:GNN:predict} 
            \algocomment{model forward pass}  \;
		Compute the loss $\mathcal{L}(q, \token_q; \theta)$ by Eq.~\eqref{eq:place:loss}; \label{line:GNN:loss} \;
			
		$\theta \leftarrow \theta-\alpha \nabla_{\theta} \mathcal{L}(q)$; \label{line:GNN:end}  
    \algocomment{update parameters of ACS model}\; 
	}
    }
	\Return{$\theta$ and $\token$};\;
\end{algorithm}
}

\begin{algorithm}[t]
	\footnotesize
    \caption{\kw{PLACEForwardPass}($\Graph$, $\token$, $q^*$)}
	\label{alg:foward}
	\DontPrintSemicolon
	\SetKwData{Up}{up}  \SetKwInOut{Input}{Input} \SetKwInOut{Output}{Output}
	\Input{input graph $\Graph$, token embedding $\token$, query $q^*$}
	\Output{the likelihood of community membership $p$}
	Construct the prompt graph $\Graph_p$ by extracting and connecting query-prompt tokens $\token_q$ of $q$; \label{line:GNN:constructgp}\;
		$\Graph_m \leftarrow \Psi(\Graph,\Graph_p)$;  \label{line:GNN:constructgm}  
            \algocomment{construct prompt-augmented graph} \;
		$p \leftarrow \rho(q, \phi_\theta(\Graph_m))$; \label{line:GNN:predict} 
            \algocomment{model forward pass}  \;
	\Return $p$\;
\end{algorithm}

\begin{algorithm}[t]
	\footnotesize
    \caption{\kw{PLACETrain}($\Graph$, $Q$, $T$, $\alpha, \beta$)}
	\label{alg:train}
	\DontPrintSemicolon
	\SetKwData{Up}{up}  \SetKwInOut{Input}{Input} \SetKwInOut{Output}{Output}
	\Input{input graph $\Graph$, training queries $Q = \{ q_1, \cdots\}$,  epochs $T$, learning rates $\alpha, \beta$}
	\Output{parameters of ACS model $\theta$ and prompt tokens $\tau$}
    Initialize parameters $\theta$ and $\tau$ \label{line:train:initial} \;
    \For { epoch $\leftarrow$ 1 to $T$ } {\label{line:train:epoch:start}
        \For {$q \in Q$}{
          Unfreeze $\tau$ and freeze $\theta$ \label{line:train:prompt:start}\;
	   $p \leftarrow$   \kw{PLACEForwardPass}($\Graph$, $\token$, $q$)  \;
            Compute the loss $\loss(q, \token_q; \theta)$ by Eq.~\eqref{eq:place:loss} \label{line:train:prompt:loss} \;
			
		$\tau_q \leftarrow \tau_q -\beta \nabla_{\tau_q} \loss(q, \token_q; \theta)$ \; \label{line:train:prompt:end}  
            Freeze $\tau$ and unfreeze $\theta$ \label{line:train:gnn:start} \;
            $p \leftarrow$   \kw{PLACEForwardPass}($\Graph$, $\token$, $q$) \;
            Compute the loss $\loss(q, \token_q; \theta)$ by Eq.~\eqref{eq:place:loss} \label{line:train:gnn:loss} \;
			
		$\theta \leftarrow \theta-\alpha \nabla_{\theta} \loss(q, \token_q; \theta)$ \; \label{line:train:epoch:end}  
            
          }
    }
	\Return $\theta$, $\tau$\;
\end{algorithm}

\subsection{Scalable to Large Graph}
\label{sec:training:scale}
Considering existing learning-based ACS approaches struggle to scale up to large graphs in limited GPU memory, we further equip \PLACE with scalable training and inference strategies. 
The training strategy is inspired by the mini-batch training of GNNs~\cite{DBLP:conf/sc/MdMMMGHKAA21, DBLP:conf/iclr/ZengZSKP20, zheng2020distdgl}, which uses a small shard of the original graph as a batch. 
In addition, the inference adopts a divide-\&-conquer strategy that processes multiple shards respectively and then merges the corresponding predictions, where the inference of shards can be sequential or parallel. 
Specifically, we partition an input large graph $\Graph$ into $s$ subgraphs (shards) $\mathbi{S} = \{\Subgraph_1, \cdots, \Subgraph_s\}$ by METIS~\cite{karypis1997metis} offline.  As METIS itself   minimizes the edge cut of the graph, the partition relieves the loss of the community structures in each shard to the largest extent. 
However, for a query $q$, its community structure, i.e., one large community or overlapping communities may also spread in multiple shards. 
Therefore, we define the query-route graph $\Subgraph_q$ that incorporates query node and involving edges to infer the community of $q$ in a shard $\Subgraph_i$.

\comment{
\begin{algorithm}[t]
	\small
    \caption{Scaling PLACE for Large Graphs}
	\label{alg:scale}
	\DontPrintSemicolon
	\SetKwData{Up}{up}  \SetKwInOut{Input}{Input} \SetKwInOut{Output}{Output}
	\Input{large data graph $\Graph$, training queries $Q=(q_1,
    \dots)$, test queries $Q^*=(q_1^*,\dots)$}
	\Output{prediction result $\hat{l}_{q^*}$ for query ${q^*} \in Q^*$}
	\SetKwFunction{Emit}{Emit}
	\SetKwFunction{Check}{Check}
        Partition $\Graph$ into $s$ subgraphs using METIS Algorithm and construct graph list $\mathcal{S}=\{\Subgraph_1,\Subgraph _2,\dots,\Subgraph_s\}$;\label{line:partition}\;
       \For {$epoch \leftarrow1 \text{ to } T$}
       { \label{line:epoch:start1}
       \For { $q \in  Q$ } 
		{ \label{line:query:start} 
            Randomly sample $s_t$ subgraphs $\mathcal{S}_t$ from $\mathcal{S}$;\label{line:query:sample}\;                      
            \For { $\Subgraph \in  \mathcal{S}_t$  }                { \label{line:sub:start}  
            Alternating Training using $\Subgraph$;\label{line:query:altertraining} \;
	}}}
    
        Take a query $q^*\in  Q^*$ for test and initial an empty $\hat{l}_{q*}$; \;
        \For { $\Subgraph \in  \mathcal{S}$ } 
		{ \label{line:subtest:start}  
            Construct the prompt graph $\Subgraph_p$ by extracting and connecting query-prompt tokens $\token_{q^*}$ of $q^*$;\label{line:subtest:constructgp}\;
		$\Subgraph_m\leftarrow\Psi(\Subgraph,\Subgraph_p)$;  \label{line:subtest:constructgm}  \;
		$\hat{l}(\Subgraph) \leftarrow \phi_\theta(\Subgraph_m)$; \label{line:subtest:predict} 
            \algocomment{model prediction for the subgraph}  \;
            Add $\hat{l}(\Subgraph)$ to $\hat{l}_{q*}$;\label{line:subtest:merge}  \algocomment{merge the result}\;
	}
    \Return $\hat{l}_{q*}$;
\end{algorithm}
}

\begin{algorithm}[t]
	\footnotesize
    \caption{\kw{PLACETrainScaleUp}($\Graph$, $Q$, $T$, $\alpha, \beta$)}
	\label{alg:train:scaleup}
	\DontPrintSemicolon
	\SetKwData{Up}{up}  \SetKwInOut{Input}{Input} \SetKwInOut{Output}{Output}
	\Input{input graph $\Graph$, training queries $Q = \{ q_1, \cdots\}$,  epochs $T$, learning rates $\alpha, \beta$}
	\Output{parameters of ACS model $\theta$ and prompt tokens $\tau$}
    Initialize parameters $\theta$ and $\tau$ \;
    {\color{blue} Partition $\Graph$ into $s$ shards $\mathbi{S} = \{\Subgraph_1, \cdots, \Subgraph_s \}$ using METIS} \label{line:train:scaleup:partition}\;
    \For { epoch $\leftarrow$ 1 to $T$ }{ 
        \For {$q \in Q$}{
            {\color{blue}Sample a shard $\Subgraph_i$ from $\mathbi{S}$} \label{line:train:scaleup:sample} \;
          {\color{blue}Construct $\Subgraph_q$ by $\Subgraph_i$ and $q$} \label{line:train:scaleup:construct}\;
          Unfreeze $\tau$ and freeze $\theta$ \;
	   $p \leftarrow$   \kw{PLACEForwardPass}({\color{blue}$\Subgraph_q$}, $\token$, $q$) \label{line:train:scaleup:forward1} \;
            Compute the loss $\loss(q, \token_q; \theta)$ by Eq.~\eqref{eq:place:loss}  \;
			
		$\tau_q \leftarrow \tau_q -\beta \nabla_{\tau_q} \loss(q, \token_q; \theta)$ \;   
            Freeze $\tau$ and unfreeze $\theta$ \;
            $p \leftarrow$   \kw{PLACEForwardPass}({\color{blue}$\Subgraph_q$}, $\token$, $q$) \label{line:train:scaleup:forward2} \;
            Compute the loss $\loss(q, \token_q; \theta)$ by Eq.~\eqref{eq:place:loss}  \;
			
		$\theta \leftarrow \theta-\alpha \nabla_{\theta} \loss(q, \token_q; \theta)$ \; \label{line:GNN:end}  
            
          }
    }
	\Return $\theta$, $\tau$\;
\end{algorithm}


\begin{definition}(Query-Route Subgraph) Given a query $q = (\VSet_q, \ASet_q)$, a shard $\Subgraph_i = (\VSet_i, \ESet_i, \ASet)$ of graph $\Graph = (\VSet, \ESet, \ASet)$, the query-route subgraph of $\Subgraph_i$ by $q$ is a graph $\Subgraph_q$ with node set $ \VSet_i \cup \VSet_q$ and edge set 
   $\ESet_i \cup \{(v, u) \mid v \in \VSet_i, u \in \VSet_q, (u, v) \in \ESet \}$.
\end{definition}

The fusion of only one-order connections of $\VSet_q$ into $\Subgraph_i$ enhances the query-centric proximity of shards, while avoiding shard expansion. And the insertion of the query prompt graph into $\Subgraph_q$ will further enrich the query node and attribute information. It is worth mentioning that the METIS partitioning is a one-off step, where the resulting shards $\mathbi{S}$ are shared by all the training and test queries. 


\stitle{Scalable training.} Training on a large graph $\Graph$ has three differences from the original alternative training algorithm, which are highlighted in Algorithm~\ref{alg:train:scaleup}. First, as a preprocessing step, the graph is partitioned by METIS in line~\ref{line:train:scaleup:partition}. Second, a shard $\Subgraph_i$ is sampled from $\mathbi{S}$ for training (line~\ref{line:train:scaleup:sample}). 
Third, we construct the query-route graph $\Subgraph_q$ (line~\ref{line:train:scaleup:construct}) and use $\Subgraph_q$ instead of the complete graph $\Graph$ for model inference (line~\ref{line:train:scaleup:forward1} \& line~\ref{line:train:scaleup:forward2}). 
The sampling on shards helps to improve the generalization of the model, while the offline graph partition reduces the online graph sampling in typical mini-batch training of GNNs. 

\stitle{Divide-\&-Conquer Inference.}
In the inference stage, the inference is performed on the query-route graph of each shard individually, and finally merges the prediction of each shard $\mathcal{S}_i$, into the complete prediction of the original graph $\Graph$. The conquered inference process can be executed sequentially and in parallel, depending on the hardware resources. 


\comment{
\begin{algorithm}[t]
	\small
    \caption{\PLACE Alternating Training Process}
	\label{alg:training}
	\DontPrintSemicolon
	\SetKwData{Up}{up}  \SetKwInOut{Input}{Input} \SetKwInOut{Output}{Output}
	\Input{input graph $\Graph=(\VSet,\ESet,\ASet)$, query $q=(\VSet_q,\ASet_q)$, number of epochs $T$}
	\Output{parameters of GNN encoder $\phi_\theta$ and prompt tokens $\token$}
	\SetKwFunction{Emit}{Emit}
	\SetKwFunction{Check}{Check}
        Initialize GNN encoder $\phi_\theta$ and prompt tokens $\token$;\label{line:encoder:initial} \;
        \For{$epoch \leftarrow 1$ to $T$}{\label{line:promptepoch:start}
        Unfreeze the parameters of prompt tokens and freeze the parameter $\theta$ of GNN encoder;  \label{line:prompt:freeze} \;
		\For { $q=(\VSet_q, \ASet_q) \in Q$ } 
		{ \label{line:prompt:start}  
            Construct $\Graph_p$ by choosing query-prompt tokens $\token_q$ according to $q=(\VSet_q, \ASet_q)$;\label{line:prompt:constructgp}\;
		$\Graph_m=\Psi(\Graph,\Graph_p)$;  \label{line:prompt:constructgm}  
            \algocomment{construct prompt-augmented graph} \;
		$H_q \leftarrow \phi_\theta(\Graph_m)$; \label{line:prompt:GNN} 
            \algocomment{compute query-level embedding}  \;
            $\hat{l_q} \leftarrow \varphi(H_q)$; \label{line:prompt:predict}\algocomment{predict by decoder}\;
		Compute the prompt Loss $\mathcal{L}(q)$ by $\hat{l_q}$ and $l_q$; \label{line:prompt:loss} \;
			
		$\token_q \leftarrow \token_q-\alpha \nabla_{\token_q} \mathcal{L}(q)$; \label{line:prompt:end}  
    \algocomment{update parameters of prompt tokens}\; 
	}
    Freeze the parameters of prompt tokens and unfreeze the parameter $\theta$ of GNN encoder;  \label{line:GNN:unfreeze} \;
    \For { $q=(\VSet_q, \ASet_q) \in Q$ } 
		{ \label{line:GNN:start} 
            Extract prompt tokens $\token_q$ according to $q=(\VSet_q, \ASet_q)$ and construct prompt graph; \label{line:GNN:constructgp}\;
		$\Graph_m=\Psi(\Graph,\Graph_p)$;  \label{line:GNN:constructgm}  
            \algocomment{construct prompt-augmented graph} \;
		$H_q \leftarrow \phi_\theta(\Graph_m)$; \label{line:GNN:GNN} 
            \algocomment{compute query-level embedding}  \;
            $\hat{l_q} \leftarrow \varphi(H_q)$; \label{line:GNN:predict}\algocomment{predict by decoder}\;
		Compute the Loss $\mathcal{L}(q)$ by $\hat{l_q}$ and $l_q$; \label{line:GNN:loss} \;
			
		$\theta \leftarrow \theta-\alpha \nabla_{\theta} \mathcal{L}(q)$; \label{line:GNN:end}  
    \algocomment{update parameters of GNN encoder}\; 
	}
    }
	\Return{$\theta$ and $\token$};\;
\end{algorithm}
}

\label{subsec:complexity}
\stitle{Complexity Analysis.}
We analyze the time complexity of \PLACE. The $K$ layer GNN encoder has a complexity of $\mathcal{O}(K |\VSet|^2 d)$ in a dense graph and $\mathcal{O}(K (|\ESet| + |\VSet|) d)$ in a sparse graph, where $d$ is the number of hidden units of GNN. 
Given that the input graph is the prompt-augmented graph $\Graph_m = (\VSet_m, \ESet_m)$. We have $|\VSet_m| = |\VSet| + |\tau_q|$, and $|\ESet_m|$ is bounded by $(|\token_q||\VSet|+|\token_q|^2 +|\ESet|)$ in the worst case where each node in the original graph possesses all query attributes and all the prompt tokens are fully connected. Since any query nodes and query attributes are of constant size, $\mathcal{O}(|\VSet_m|) = |\VSet|$ and $\mathcal{O}(|\ESet_m|) = |\VSet| + |\ESet|$.
This indicates that taking $\Graph_m$ as the model input will not increase the complexity. In addition to the complexity of the inner-product decoder, $\mathcal{O}(|\VSet|d)$, the overall complexity of \PLACE remains $\mathcal{O}(K |\VSet|^2 d)$ and $\mathcal{O}(K (|\ESet| + |\VSet|) d)$ for dense graphs and sparse graphs, respectively.

\section{Why Graph Prompt Work for ACS?}
\label{sec:analysis}

The essence of tuning the prompt is to enrich the contextualized input. For classical graph learning tasks, existing literature, All-in-One~\cite{DBLP:conf/kdd/SunCLLG23}, analyzes that we can learn an optimal prompt graph $\Graph_p^*$ for a graph $\Graph$. The insertion of the prompt graph imitates any graph-level transformation $\bm{g}$, such as modifying node features, adding or removing nodes and edges. This flexible, learnable transformation enables GNN models to achieve superior accuracy in graph learning tasks.
When extending to ACS tasks, given a graph $\Graph$ with adjacency matrix $A$ and node feature matrix $X$, \PLACE inserts a query-prompt graph $\Graph_p$ for a query $q$, which is approximately equivalent to applying a query-specific graph transformation $\bm{g}$ as follows:
\begin{align}
\phi\left(\Psi(\Graph,\Graph_p)\right)=\phi\left(\bm{g}(A,X,q)\right)+O^*_{\phi q}~,\label{eq:cs} 
\end{align}
where $\phi$ is the GNN encoder of \PLACE, and $O^*_{\phi q}$ denotes the error bound between the transformed graph and the prompt-augmented graph w.r.t. their representations from the GNN encoder.

A latest study~\cite{wang2024does}
proves a theoretical upper bound on the error between the representations learned from a prompt-augmented graph and those from an ideally structured graph. 
Specifically, the error bound of our task is shown below:
\begin{align}
O^*_{\phi q}=\|\phi\left(\Psi(\Graph,\Graph_p)\right)-\phi\left(\bm{g}(A,X,q)\right)\| \nonumber \\
\leq \sin{(\Phi/2)} \|\phi\left(\bm{g}(A,X,q)\right)\|. \nonumber 
\end{align}
Intuitively, the error $O^*_{\phi q}$ is bounded by the model  ($\sin(\Phi/2)$) and the graph data $\|\phi\left(\bm{g}(A,X,q)\right)\|$. 
In practical scenarios, since the magnitudes of node features and the size of the node set are bounded, $\|\phi(\bm{g}(A,X,q))\|$ is upper bounded by a constant, ensuring that the overall representation error $O^*_{\phi q}$ is also bounded. This demonstrates that the prompt mechanism can approximate the effect of optimal graph transformations within a certain error range. In the rest of this section, we show an empirical study to further verify the error bound and discuss how existing ACS approaches manipulate the original graph, which can be regarded as `fixed' prompting strategies.


\begin{figure}[]
    \centering
    \includegraphics[width=0.3\textwidth]{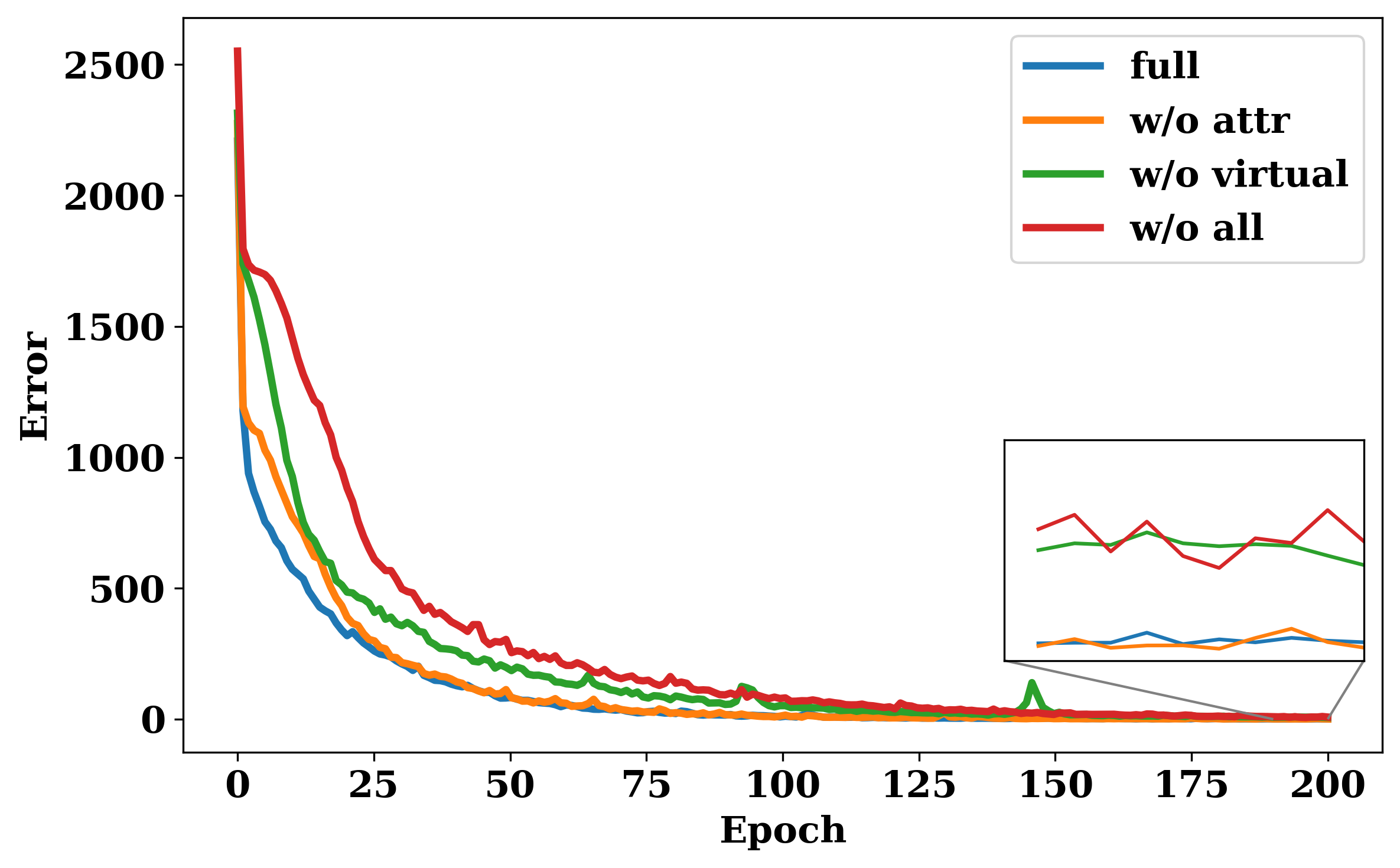}
    \hspace{-2ex}
    \caption{Training Errors with Different Prompts}
    \label{fig:error}
    \hspace{-2ex}
\end{figure}

\label{sec:exp:emp}
\stitle{Empirical Study on Training Errors.}
As discussed above, the prompt of \PLACE can effectively manipulate the original graph within a specific error bound. We conduct an empirical analysis to compare the error of different prompt token configurations for GNN models, as shown in Fig.~\ref{fig:error}. Specifically, we evaluate the training error between the ground truth communities and the outputs of the GNN under four configurations: full prompt (\PLACE), w/o attribute  tokens, w/o virtual node tokens, and w/o all (vanilla GNN). The results indicate that the error (loss) of \PLACE converges fastest during the entire training process, while the vanilla GNN consistently exhibits the highest error, which reflect that \PLACE is more powerful to manipulate graph transformations for ACS.
Regarding the converged models, 
the errors for the full prompt and w/o attribute tokens are lower than those w/o all and w/o virtual node  tokens. This observation is also consistent with the results in our ablation study~(\cref{sec:exp:ablation}), further confirming that virtual prompt tokens play a more critical role than attribute tokens.


\stitle{Connection to Existing Approaches.} 
We discuss how existing ACS approaches manipulate the original graph, which can be regarded as `fixed' prompting strategies.
Existing learning-based ACS approaches have intrinsic connections to graph prompting, which manipulates original graphs before or during the GNN processing. Here, we discuss four representative approaches with GNN as their model backbone. 
Specifically, \ICSGNN and \ALICE extract candidate subgraphs using heuristic strategies, while \IACS and \AQDGNN enhance node embeddings and query embeddings through specialized modules:
\ding{182} For a query node, \ICSGNN~\cite{ICSGNN, DBLP:journals/vldb/ChenGC23} iteratively extracts a candidate subgraph by BFS as an initial coarse community and then trains a GNN model on the subgraph. 
\ding{183} \ALICE~\cite{ALICE} also extracts a candidate subgraph near query nodes by maximizing modularity from a node-attribute bipartite graph. Subsequently, 
a \GIN~\cite{DBLP:conf/iclr/XuHLJ19} model generates node embeddings for the candidate subgraph.
\ding{184} \IACS~\cite{IACS} introduces a pre-trained attributed embedding that aligns heterogeneous attribute sets across different graphs. In addition, the embedding of query $q$, formulated by the fusion of query node embedding and query attribute embedding, serves as a special prompt. 
\ding{185} Beyond a GNN Graph Encoder, \AQDGNN~\cite{AQDGNN} employs two additional GNN encoders, a Query Encoder and an Attribute Encoder, to encode the structural information from query nodes and the attribute information from query attributes respectively. The embeddings of the two encoders are fused into the node embedding from the Graph Encoder to predict the final result. Through this enhancement of node embeddings, these two encoders play a similar role as our virtual node tokens and attribute tokens. 

Compared with these GNN-based ACS approaches, \PLACE provides a more flexible strategy for  manipulating the nodes, edges, and features in the input graph by inserting a learnable, query-specific query-prompt graph.
The prompt influences node embeddings, containing the query embedding, via the message propagation of GNN, and is jointly optimized together with the GNN model. 
\section{Experimental Studies}
\label{sec:exp}

We introduce the experimental setup in \cref{sec:exp:setup} and test our \PLACE in the following facets:
\ding{172} Compare the effectiveness of \PLACE on three types of ACS queries with baselines (\cref{sec:exp:compresult}).
\ding{173} Analyze the efficiency of \PLACE (\cref{sec:exp:efficiency}).
\ding{174} Conduct an ablation study to investigate the impact of prompts (\cref{sec:exp:ablation}).
\ding{175} Investigate the sensitivity of \PLACE regarding different configurations (\cref{sec:exp:parameteranalysis}).
\ding{176} Study the transferability of \PLACE (\cref{sec:exp:transferability}).
\ding{177} Conduct a visualized case study of prompt-augmented graphs (\cref{sec:exp:casestudy}).

\begin{table}[t]
	\caption{The Profiles of Dataset}
	\label{tab:dataset}
	\scriptsize
    \centering
    \resizebox{0.4\textwidth}{!}{
		\begin{tabular}{c|r r r r}
			\toprule
      {\textbf{Dataset}}  & \textbf{$|\VSet|$} &\textbf{ $|\ESet|$}  & \textbf{$|\ASet|$} & \textbf{$|\community|$}  
      \\\midrule       
            {\Cornell}~\cite{pei2020geom}   & 183 & 280  &1,703   &5\\
            {\Texas}~\cite{pei2020geom}   & 183 & 295  &1,703   &5\\
            {\Washington}~\cite{pei2020geom}   & 215 & 402  &1,703   &5\\
            {\Wisconsin}~\cite{pei2020geom}   & 251 & 466  &1,703   &5\\
                {\Cora}~\cite{yang2016revisiting}   & 2,708 & 5,429  & 1,433  &7\\ 
			{\Citeseer}~\cite{yang2016revisiting}  & 3,327 & 4,732 & 3,703 & 6 \\ 
            {\Reddit}~\cite{hamilton2017inductive}   &232,965 &114,615,892 &1,164 & 50\\  
            {\Product}~\cite{chiang2019cluster}    &2,449,029 &61,859,140   &12,245 &47 \\  
		 {\Orkut}~\cite{leskovec2012learning}    &3,072,441 &117,186,083   &15,362 &5,000 \\	
            \bottomrule
	\end{tabular}}
\end{table}

\begin{table*}[t]
\caption{Overall Performance on \AFC(\%)}
\label{tab:effctiveness}
\centering
\scriptsize
\resizebox{0.9\textwidth}{!}{
\begin{tabular}{c|ccc|ccc|ccc}
\toprule
\rowcolor{gray!20}
 & \Pre           & \Rec          & \Fone              & \Pre           & \Rec          & \Fone  & \Pre           & \Rec          & \Fone                 \\
\cmidrule{2-10}
{Method}                                  & \multicolumn{3}{c|}{\Cora}   & \multicolumn{3}{c|}{\Citeseer}   & \multicolumn{3}{c}{\Cornell}                    \\ 

 \ATC & $23.86_{\pm2.64}$ & $1.99_{\pm0.01}$ & $3.86_{\pm1.38}$  & $21.47_{\pm3.71}$ & $0.68_{\pm0.25}$ & $1.32_{\pm0.47}$ & $19.13_{\pm3.72}$ & $2.13_{\pm0.94}$ & $3.79_{\pm1.59}$ \\
\ACQ & $69.79_{\pm0.30}$ & $5.92_{\pm0.11}$ & $11.39_{\pm0.87}$ & $44.91_{\pm0.13}$ & $3.19_{\pm0.01}$ & $5.97_{\pm0.05}$ & $20.84_{\pm5.89}$ & $1.28_{\pm0.57}$ & $2.40_{\pm1.06}$ \\
 \AQDGNN & $49.54_{\pm0.50}$ & $97.66_{\pm1.18}$ & $65.83_{\pm0.70}$ & $50.88_{\pm0.29}$ & $98.68_{\pm0.52}$ & $67.14_{\pm0.18}$ & $53.99_{\pm3.34}$ & $97.44_{\pm1.74}$ & $70.07_{\pm2.80}$ \\
\IACS   & $73.93_{\pm6.04}$ & $78.51_{\pm2.80}$ & $76.15_{\pm2.59}$ & $76.60_{\pm1.47}$ & $75.49_{\pm1.13}$ & $76.04_{\pm1.16}$ & $87.81_{\pm1.36}$ & $84.23_{\pm1.87}$ & $85.98_{\pm1.49}$ \\
\PLACE  &$87.34_{\pm1.57}$ & $95.57_{\pm4.87}$ & $\textbf{91.21}_{\pm2.05}$ & $85.05_{\pm1.39}$ & $94.62_{\pm3.09}$ & $\textbf{89.54}_{\pm0.75}$ & $82.46_{\pm3.26}$ & $98.60_{\pm0.81}$ & $\textbf{90.17}_{\pm1.89}$\\ 
\midrule
{Method}                                  & \multicolumn{3}{c|}{\Texas}   & \multicolumn{3}{c|}{\Washington}   & \multicolumn{3}{c}{\Wisconsin}                    \\ 
 \ATC & $27.53_{\pm2.88}$  & $1.46_{\pm0.52}$ & $2.76_{\pm0.91}$ & $28.27_{\pm1.99}$ & $2.01_{\pm0.69}$ & $3.73_{\pm1.19}$ & $27.88_{\pm3.03}$ & $1.50_{\pm0.42}$ & $2.84_{\pm0.76}$ \\
\ACQ & $16.92_{\pm19.83}$ & $1.28_{\pm0.04}$ & $2.70_{\pm0.42}$ & $29.57_{\pm3.06}$ & $1.73_{\pm0.14}$ & $3.27_{\pm0.27}$ & $31.18_{\pm2.50}$ & $1.19_{\pm0.11}$ & $2.29_{\pm0.21}$ \\
 \AQDGNN & $54.68_{\pm0.98}$ & $98.05_{\pm0.57}$ & $72.84_{\pm1.04}$ & $50.51_{\pm1.17}$ & $96.31_{\pm0.56}$ & $70.22_{\pm0.83}$ & $49.71_{\pm0.85}$ & $97.92_{\pm0.51}$ & $69.97_{\pm1.11}$ \\
\IACS   & $93.05_{\pm0.85}$ & $92.35_{\pm0.27}$ & $92.70_{\pm0.44}$ & $91.12_{\pm0.61}$ & $86.29_{\pm1.04}$ & $88.64_{\pm0.78}$ & $91.57_{\pm1.69}$ & $90.13_{\pm1.73}$ & $90.84_{\pm1.67}$ \\
\PLACE  &$89.44_{\pm4.41}$ & $98.07_{\pm0.98}$ & $\textbf{93.55}_{\pm2.80}$ & $92.44_{\pm1.77}$ & $99.60_{\pm0.28}$ & $\textbf{94.90}_{\pm0.85}$ & $92.56_{\pm1.86}$ & $99.64_{\pm0.23}$ & $\textbf{95.96}_{\pm0.91}$ \\  \midrule
{Method}                                  & \multicolumn{3}{c|}{\Reddit}   & \multicolumn{3}{c|}{\Product}   & \multicolumn{3}{c}{\Orkut}   \\ 
\ATC                                  & $83.29_{\pm0.63}$	&$39.08_{\pm0.11}$	&$53.19_{\pm0.04}$       &$36.91_{\pm4.37}$	&$1.75_{\pm0.35}$	&$2.94_{\pm0.09}$      &$67.77_{\pm1.32}$	&$23.42_{\pm0.07}$	&$34.81_{\pm0.02}$  \\
\ACQ                                 &$99.49_{\pm0.69}$	&$22.17_{\pm0.24}$	&$38.26_{\pm2.46}$                  & $40.91_{\pm0.13}$	&$2.81_{\pm0.26}$	&$3.44_{\pm0.62}$        & $75.00_{\pm0.09}$	&$3.44_{\pm0.69}$	&$6.59_{\pm0.15}$     \\
\AQDGNN & \OOM               &  \OOM               & \OOM               & \OOM              & \OOM           & \OOM               & \OOM               & \OOM               & \OOM               \\
\IACS  &$92.94_{\pm1.24}$	&$95.33_{\pm0.75}$	&$\textbf{94.12}_{\pm0.89}$ &$80.07_{\pm0.10}$	&$90.86_{\pm3.36}$	&$85.11_{\pm1.42}$               & \OOM               & \OOM               & \OOM               \\
\PLACE &$89.87_{\pm0.68}$	&$92.52_{\pm0.69}$	&$91.17_{\pm0.02}$	&$91.64_{\pm7.29}$	&$84.89_{\pm4.72}$	& $\textbf{87.87}_{\pm1.11}$  &$98.39_{\pm0.28}$ &$84.48_{\pm1.30}$  &$\textbf{92.05}_{\pm0.52}$ \\  
\bottomrule
\end{tabular}}
\end{table*}

\begin{table*}[t]
\caption{Overall Performance on \AFN(\%)}
\label{tab:effctiveness2}
\centering
\scriptsize
\resizebox{0.9\textwidth}{!}{
\begin{tabular}{c|ccc|ccc|ccc}
\toprule
\rowcolor{gray!20}
 & \Pre           & \Rec          & \Fone              & \Pre           & \Rec          & \Fone  & \Pre           & \Rec          & \Fone                 \\
\cmidrule{2-10}
{Method}                                  & \multicolumn{3}{c|}{\Cora}   & \multicolumn{3}{c|}{\Citeseer}   & \multicolumn{3}{c}{\Cornell}                    \\
 \ATC & $18.11_{\pm4.62}$ & $0.61_{\pm0.19}$ & $1.15_{\pm0.37}$ & $23.17_{\pm18.80}$ & $0.72_{\pm0.68}$ & $1.39_{\pm1.31}$ & $22.56_{\pm7.31}$  & $2.61_{\pm0.84}$ & $4.66_{\pm1.47}$ \\
\ACQ & $16.82_{\pm0.25}$ & $1.57_{\pm0.61}$ & $2.17_{\pm0.04}$ & $20.10_{\pm3.45}$  & $0.21_{\pm0.04}$ & $0.41_{\pm0.08}$ & $31.25_{\pm14.74}$ & $1.44_{\pm0.30}$ & $2.74_{\pm0.61}$ \\
 \AQDGNN & $49.51_{\pm0.50}$ & $98.61_{\pm0.80}$ & $65.88_{\pm0.70}$ & $50.58_{\pm0.75}$ & $98.01_{\pm0.83}$ & $67.01_{\pm0.36}$ & $50.14_{\pm0.84}$ & $97.43_{\pm1.02}$ & $70.29_{\pm0.49}$ \\
\IACS   & $72.12_{\pm4.31}$ & $79.84_{\pm2.31}$ & $75.79_{\pm2.35}$ & $75.69_{\pm1.02}$ & $74.46_{\pm1.50}$ & $75.07_{\pm0.91}$ & $87.03_{\pm1.58}$  & $78.53_{\pm2.81}$ & $82.56_{\pm2.14}$ \\
\PLACE  & $85.41_{\pm2.28}$ & $96.74_{\pm2.63}$ & $\textbf{90.24}_{\pm1.07}$ & $83.55_{\pm1.47}$  & $92.51_{\pm3.64}$ & $\textbf{87.79}_{\pm2.24}$ & $76.40_{\pm6.65}$  & $98.38_{\pm1.43}$ & $\textbf{85.86}_{\pm3.65}$   \\ 
\midrule
{Method}                                  & \multicolumn{3}{c|}{\Texas}   & \multicolumn{3}{c|}{\Washington}   & \multicolumn{3}{c}{\Wisconsin}                    \\
 \ATC & $28.35_{\pm2.49}$ & $1.78_{\pm0.72}$ & $3.32_{\pm1.27}$ & $25.39_{\pm4.96}$ & $1.71_{\pm0.39}$ & $3.19_{\pm0.67}$ & $27.34_{\pm3.59}$ & $1.57_{\pm0.59}$ & $2.95_{\pm1.06}$ \\
\ACQ & $28.78_{\pm5.08}$ & $1.17_{\pm0.18}$ & $2.25_{\pm0.36}$ & $29.33_{\pm1.36}$ & $1.80_{\pm0.01}$ & $3.39_{\pm0.01}$ & $34.41_{\pm2.91}$ & $1.29_{\pm0.12}$ & $2.48_{\pm0.23}$ \\
 \AQDGNN   & $55.96_{\pm0.88}$ & $98.60_{\pm0.62}$ & $72.35_{\pm0.91}$ & $51.66_{\pm2.13}$ & $98.22_{\pm1.00}$ & $68.31_{\pm2.02}$ & $50.02_{\pm0.79}$ & $97.80_{\pm0.63}$ & $69.32_{\pm0.46}$ \\
\IACS   & $89.69_{\pm0.35}$ & $91.62_{\pm0.31}$ & $90.64_{\pm0.15}$ & $89.33_{\pm0.31}$ & $86.88_{\pm0.59}$ & $88.09_{\pm0.14}$ & $87.47_{\pm1.69}$  & $86.39_{\pm1.73}$ & $86.92_{\pm1.67}$ \\
\PLACE  & $91.33_{\pm2.68}$ & $99.68_{\pm0.28}$ & $\textbf{95.31}_{\pm1.58}$ & $86.00_{\pm2.62}$  & $99.24_{\pm0.27}$ & $\textbf{92.27}_{\pm1.60}$ & $89.32_{\pm3.35}$  & $99.37_{\pm0.28}$ & $\textbf{94.06}_{\pm1.89}$\\  \midrule
{Method}                                  & \multicolumn{3}{c|}{\Reddit}   & \multicolumn{3}{c|}{\Product}   & \multicolumn{3}{c}{\Orkut}  \\ 
\ATC                                  & $91.38_{\pm12.07}$	&$32.68_{\pm9.15}$	&$47.35_{\pm8.22}$      &$33.63_{\pm0.53}$	&$1.78_{\pm0.31}$	&$3.05_{\pm0.08}$        &$30.89_{\pm0.18}$	&$18.94_{\pm0.50}$	&$23.48_{\pm0.98}$  \\
\ACQ                  &$99.46_{\pm0.64}$	&$26.61_{\pm0.56}$	&$41.77_{\pm0.33}$      &$40.05_{\pm0.07}$	&$1.07_{\pm0.04}$	&$2.32_{\pm0.02}$     &$90.00_{\pm0.18}$	&$1.44_{\pm0.47}$	&$2.84_{\pm1.00}$ \\
\AQDGNN & \OOM              & \OOM               & \OOM               & \OOM               & \OOM               & \OOM               & \OOM                & \OOM               & \OOM               \\
\IACS   &$83.59_{\pm2.40}$	&$92.76_{\pm0.61}$	&$\textbf{87.94}_{\pm1.25}$ &$84.24_{\pm3.44}$	&$91.56_{\pm1.02}$	& $\textbf{87.72}_{\pm1.40}$                & \OOM                & \OOM               & \OOM               \\
\PLACE  & $96.27_{\pm0.10}$	& $71.70_{\pm3.61}$ &	$82.16_{\pm2.34}$	& $85.08_{\pm3.82}$ &	$84.15_{\pm5.46}$	& {$84.45_{\pm1.19}$} & $99.91_{\pm3.84}$  & $86.41_{\pm2.64}$ & $\textbf{91.41}_{\pm1.11}$\\  
\bottomrule
\end{tabular}}
\end{table*}

\begin{table*}[t]
\caption{Overall Performance on \EMA(\%)}
\label{tab:effctiveness3}
\centering
\scriptsize
\resizebox{0.9\textwidth}{!}{
\begin{tabular}{c|ccc|ccc|ccc}
\toprule
\rowcolor{gray!20}
 & \Pre           & \Rec          & \Fone              & \Pre           & \Rec          & \Fone  & \Pre           & \Rec          & \Fone                 \\
\cmidrule{2-10}
{Method}                                  & \multicolumn{3}{c|}{\Cora}   & \multicolumn{3}{c|}{\Citeseer}   & \multicolumn{3}{c}{\Cornell}                \\ 

 \CTC       & $18.27_{\pm1.73}$  & $0.30_{\pm0.01}$  & $0.59_{\pm0.02}$  & $15.98_{\pm0.04}$ & $0.39_{\pm0.01}$  & $0.88_{\pm0.18}$  & $22.22_{\pm6.46}$ & $2.42_{\pm0.14}$  & $4.34_{\pm0.09}$   \\
 \Transzero & $48.08_{\pm1.34}$ & $68.11_{\pm1.85}$ & $56.36_{\pm1.51}$ & $44.34_{\pm0.16}$ & $45.19_{\pm0.62}$ & $44.76_{\pm0.34}$ & $29.26_{\pm0.21}$ & $44.71_{\pm0.56}$ & $35.37_{\pm0.19}$ \\
\QDGNN     & $50.79_{\pm0.50}$ & $98.56_{\pm0.67}$ & $67.03_{\pm0.53}$ & $50.99_{\pm0.25}$ & $97.55_{\pm0.51}$ & $66.97_{\pm0.24}$ & $50.42_{\pm1.78}$ & $97.81_{\pm1.07}$ & $72.22_{\pm0.80}$ \\
\IACS      & $64.64_{\pm0.38}$ & $85.12_{\pm0.68}$ & $73.48_{\pm0.50}$ & $69.36_{\pm0.19}$ & $79.82_{\pm0.61}$ & $74.22_{\pm0.24}$ & $66.81_{\pm1.91}$ & $96.51_{\pm0.43}$ & $79.24_{\pm1.27}$ \\
\PLACE     & $90.55_{\pm2.42}$  & $93.74_{\pm2.35}$ & $\textbf{92.09}_{\pm1.49}$ & $86.86_{\pm2.53}$ & $92.98_{\pm3.54}$ & $\textbf{89.78}_{\pm2.17}$ & $89.42_{\pm1.99}$ & $98.74_{\pm0.64}$ & $\textbf{94.06}_{\pm1.25}$  \\ 
\midrule
{Method}                                  & \multicolumn{3}{c|}{\Texas}   & \multicolumn{3}{c|}{\Washington}   & \multicolumn{3}{c}{\Wisconsin}                     \\ 
 \CTC       & $24.89_{\pm12.81}$ & $1.55_{\pm0.75}$  & $2.92_{\pm1.42}$  & $34.98_{\pm4.73}$ & $2.50_{\pm0.56}$  & $4.67_{\pm1.02}$  & $25.89_{\pm8.96}$ & $1.49_{\pm0.55}$  & $2.81_{\pm1.04}$  \\
 \Transzero & $30.77_{\pm1.64}$ & $44.21_{\pm1.04}$ & $36.24_{\pm0.77}$ & $26.87_{\pm0.29}$ & $44.83_{\pm1.76}$ & $33.55_{\pm0.53}$ & $34.02_{\pm0.35}$ & $48.60_{\pm1.35}$ & $40.02_{\pm0.46}$ \\
\QDGNN     & $56.52_{\pm0.94}$ & $99.20_{\pm0.89}$ & $75.06_{\pm0.83}$ & $53.47_{\pm1.91}$ & $98.21_{\pm0.14}$ & $74.05_{\pm1.11}$ & $54.31_{\pm1.24}$ & $97.49_{\pm0.77}$ & $71.34_{\pm0.79}$ \\
\IACS      & $82.43_{\pm1.80}$ & $98.01_{\pm0.63}$ & $89.55_{\pm1.21}$ & $71.22_{\pm4.30}$ & $98.76_{\pm0.20}$ & $82.76_{\pm2.83}$ & $68.94_{\pm4.85}$ & $92.46_{\pm3.47}$ & $78.98_{\pm4.39}$ \\
\PLACE      & $94.18_{\pm1.95}$  & $98.77_{\pm0.90}$ & $\textbf{96.42}_{\pm1.15}$ & $92.01_{\pm1.89}$ & $97.92_{\pm1.01}$ & $\textbf{94.88}_{\pm1.26}$ & $91.88_{\pm1.75}$ & $98.18_{\pm0.62}$ & $\textbf{94.92}_{\pm1.21}$\\ \midrule
{Method}                                  & \multicolumn{3}{c|}{\Reddit}   & \multicolumn{3}{c|}{\Product}   & \multicolumn{3}{c}{\Orkut}   \\ 
\CTC                             &$90.46_{\pm0.64}$	&$4.99_{\pm0.02}$	&$9.72_{\pm0.40}$         &$26.88_{\pm0.17}$	&$0.28_{\pm0.04}$	&$7.50_{\pm3.54}$         &$99.50_{\pm0.70}$	&$0.78_{\pm0.04}$	&$1.49_{\pm0.01}$                  \\
\Transzero & $41.46_{\pm0.45}$ & $74.32_{\pm0.77}$ & $53.23_{\pm0.56}$ & $28.24_{\pm0.39}$ & $7.07_{\pm0.10}$  & $11.30_{\pm0.16}$ & $0.06_{\pm0.03}$  & $3.03_{\pm1.75}$  & $0.12_{\pm0.07}$  \\
\QDGNN      & \OOM               & \OOM               & \OOM               & \OOM               & \OOM               & \OOM              & \OOM               & \OOM               & \OOM               \\
\IACS  &$91.47_{\pm2.41}$	&$95.29_{\pm0.77}$	& $\textbf{93.34}_{\pm1.38}$ &$81.80_{\pm3.44}$	&$88.40_{\pm1.02}$	&$\textbf{84.44}_{\pm1.40}$                 & \OOM               & \OOM               & \OOM               \\
\PLACE     & $83.59_{\pm1.41}$	&$92.76_{\pm1.68}$ & $87.94_{\pm1.58}$	&$91.12_{\pm7.94}$ & $70.93_{\pm3.26}$	&{$79.63_{\pm3.66}$} & $99.84_{\pm2.24}$ &$85.86_{\pm0.66}$ & $\textbf{92.32}_{\pm1.38}$ \\  
\bottomrule
\end{tabular}}
\end{table*}

\subsection{Experimental Setup}
\label{sec:exp:setup}

\stitle{Datasets.}
We conduct experiments using nine real-world graph datasets, summarized in Table~\ref{tab:dataset}. The notations $|\ASet|$ and $|\community|$ represent the number of  distinct attributes and communities, respectively.
The first four datasets are from WebKB~\cite{pei2020geom}, consisting of web pages where nodes are the pages and edges are hyperlinks. 
These datasets contain $1,703$ unique words as attributes.
\Cora and \Citeseer are both citation networks~\cite{yang2016revisiting} where nodes represent papers and edges represent citation links. Attributes are words from a predefined dictionary. Ground-truth communities are formed by connected nodes sharing the same research topic.
To evaluate the scalability of \PLACE, we use three large graphs \Reddit, \Product and \Orkut. Since these datasets lack discrete attributes, we generate synthetic attributes following the protocol of \cite{ATC}. \Reddit~\cite{hamilton2017inductive} contains Reddit posts as nodes, with edges between nodes representing post co-comments. The communities correspond to various post categories. 
\Product~\cite{chiang2019cluster} is an Amazon co-purchasing network, where nodes represent products, and edges indicate co-purchases. The communities represent product categories. 
\Orkut~\cite{leskovec2012learning} is a social network based on user friendships, where user-defined groups serve as the ground-truth communities. 
We use binary vector representations of the attributes as the node features. 
The METIS algorithm partitions the three large graphs into shards, each containing approximately $1,000$ nodes. In one training epoch, we randomly sample $3$ shards to train \PLACE for each query.

\stitle{Queries.}
Regarding query composition, the number of query nodes, $|\VSet_q|$, ranges from 1 to 3, allowing us to evaluate the cases of both single and multiple query nodes.
For query attributes $\ASet_q$, we explore three distinct cases of $\ASet_q$ following the literature~\cite{ALICE, AQDGNN}:
\ding{172}
\emph{Attribute from community} (\AFC) aims at searching a community based on attributes and nodes of interest. We identify the top 3 frequently occurring attributes among all nodes within a community and use them as query attributes of the community.
\ding{173}
\emph{Attribute from the query node} (\AFN) is designed for node-oriented queries, where the searched community should not only include the query nodes but also possess attributes related to those query nodes. Query attributes are derived directly from the query nodes, i.e., $\ASet_q=\bigcup_{v\in \VSet_q} \ASet(v)$. 
\ding{174}
\emph{Empty query attribute} (\EMA) is for non-attributed community search, which emphasizes structural cohesiveness. Queries are free of query attributes, i.e., $\ASet_q= \emptyset$.

{For all datasets, we generate a total of $|Q|=150$ queries and split the queries into training, validation and test sets, each set containing $50$ queries. 
For the three large graphs, we utilize the total training labels for each sampled shard. For the remaining datasets, following the setup of \cite{IACS}, we randomly sample 5\% nodes from $\community_q$ and the remaining nodes as the positive and negative training labels, respectively.}


\stitle{Baselines.}
{We compare \PLACE against six baseline approaches, including three algorithmic approaches: \CTC~\cite{CTC}, \ATC~\cite{ATC}, \ACQ~\cite{ACQ}, and three learning-based approaches: \Transzero~\cite{wang2024efficient}, {{\sl QD/AQD-GNN}}~\cite{AQDGNN}, and \IACS~\cite{IACS}.}
\stitle{Implementation details.}
For the GNN encoder of \PLACE, we try \GAT~\cite{velickovic2017graph}, \GIN~\cite{DBLP:conf/iclr/XuHLJ19} and \RGCN~\cite{schlichtkrull2018modeling}, and use \RGCN as the default GNN model.
The encoder is composed of 3 \RGCN layers, each containing 128 hidden units.
The learning framework of \PLACE is built on PyTorch~\cite{pytorch} and PyTorch Geometric~\cite{torchgeo}. 
By default, we set the number of virtual node tokens as $1$, which is insensitive to the performance in our testing. 
We train both the GNN and prompt using the Adam optimizer for 200 epochs, with a learning rate of 
1e-4. Other learning-based baselines maintain their default configurations during training.
All learning-based approaches are tested on a Tesla A100 with 80GB memory. Algorithmic approaches (\CTC,\ATC and \ACQ) are evaluated on the same Linux server with 96 AMD EPYC 7413 CPUs and 512GB RAM.


\subsection{Overall Effectiveness}
\label{sec:exp:compresult}
We evaluate the overall effectiveness of \PLACE, comparing against six baselines. To ensure statistical significance, each experiment runs three times with different random seeds. We report the mean and the standard deviation in Table~\ref{tab:effctiveness}-\ref{tab:effctiveness3}, where the best \Fone is in bold. Here, `\OOM' indicates the GPU runs out-of-memory.

\subsubsection{Attributed Community Search} 
Table~\ref{tab:effctiveness} and Table \ref{tab:effctiveness2} present the overall performance on \AFC and \AFN, respectively, which demonstrate that \PLACE surpasses all baselines in 15 out of 18 cases.
The superiority of \PLACE lies in its significant improvement in Precision, while maintaining a high Recall up to $85\%$ in most cases. 
Notably, \PLACE is the only learning-based method scalable to million-scale graphs, i.e., \Orkut, highlighting its scalability. 
The algorithmic approaches, \ATC and \ACQ lead to low Recall, primarily due to their inflexible design of a two-stage search process, which restricts the discovery of promising nodes. 
{\AQDGNN also underperforms with a high recall but low precision since the model tends to make a positive prediction.} 
In contrast, \IACS achieves competitive performance, particularly on \Reddit and \Product.
As an inductive learning approach, \IACS is deployed by a different `training-adaptation-inference' workflow, which trains a model by a collection of ACS tasks and then fine-tunes the model on the test ACS task. 
In our experiments, for \Reddit and \Product, disjoint subgraphs extracted from the original graph are used as a task. For other datasets, original graphs with different queries are used as tasks.
Under this different workflow and setting, \IACS acquires powerful generalization capability by learning shared knowledge across tasks and fine-tuning on the test task, while the model faces the risk of overfitting on the test task.


\begin{figure}[t]
  \centering
  \vspace{-2ex}
  \includegraphics[width=0.25\textwidth]{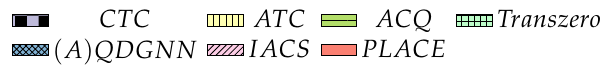}
  \resizebox{0.45\textwidth}{!}{
  \begin{tabular}[h]{c}  
  \subfigure[Training for \AFC]{
\includegraphics[width=0.22\textwidth]{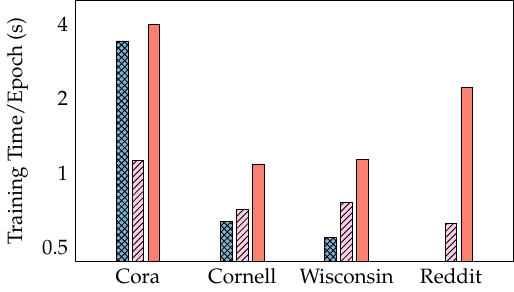}
    \label{fig:afc_training}}
   \hspace{-1ex} 
   \subfigure[Test for \AFC]{
\includegraphics[width=0.22\textwidth]{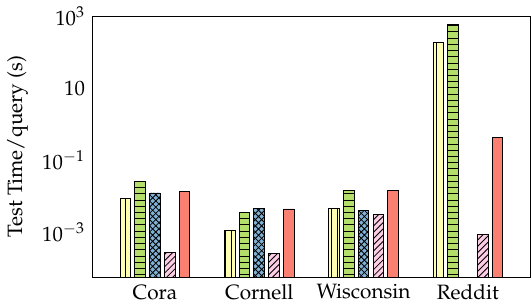}
  \label{fig:afc_test}}
  \hspace{-1ex}\\
   \subfigure[Training for \AFN]{
\includegraphics[width=0.22\textwidth]{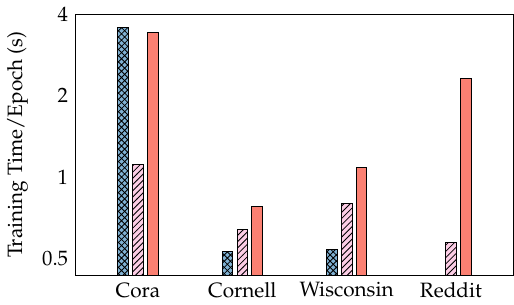}
    \label{fig:afn_training}}
   \hspace{-1ex} 
   \subfigure[Test for \AFN]{
\includegraphics[width=0.22\textwidth]{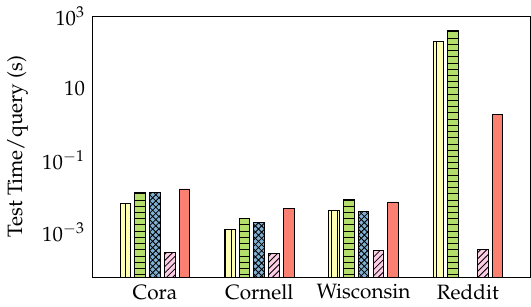}
  \label{fig:afn_test}}
  \hspace{-1ex}\\
   \subfigure[Training for \EMA]{
\includegraphics[width=0.22\textwidth]{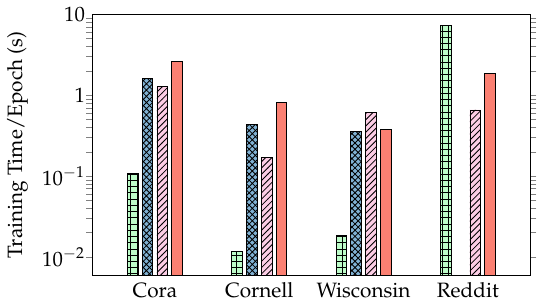}
    \label{fig:ema_training}}
   \hspace{-1ex}
  \subfigure[Test for \EMA]{
\includegraphics[width=0.22\textwidth]{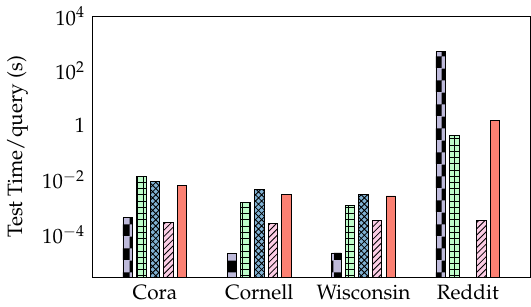}
  \label{fig:ema_test}}
  \hspace{-1ex}
  \end{tabular}
  }
   \caption{Comparison of Training and Test Time (s)}
 \label{fig:efficiency}
\end{figure}

\subsubsection{Non-attributed Community Search}
Table~\ref{tab:effctiveness3} summarizes the results on \EMA. In general, \PLACE achieves the highest \Fone score in 7 out of 9 datasets, indicating its effectiveness in non-attributed community search scenario. Notably, 
\CTC exhibits low Recall due to its struggle to flexibly identify community structure, which is consistent with the results of algorithmic approaches \ATC and \ACQ. 
\Transzero can scale to large graphs on \EMA, since it employs a Graph Transformer using subgraph-level tokenization, 
making it memory-efficient to process large-scale graphs.
However, as an unsupervised approach, \Transzero learns from node proximity self-supervision, whose performance cannot catch up with supervised learning approaches. 
The findings for \QDGNN and \IACS align similarly with those observed in the \AFC and \AFN setting.

Intuitively, \AFN is more challenging than \AFC since the query attributes of \AFN are less relevant to the communities. In the experiments, we also observe that the results in \AFC (Table~\ref{tab:effctiveness2}) are generally worse than the results in \AFN (Table~\ref{tab:effctiveness}).
In \EMA scenarios, the query focuses solely on structural cohesiveness without the constraints of attributes, which may simplify the task compared to ACS. Thereby, the results reported in Table~\ref{tab:effctiveness3} are better in general.

\subsection{Efficiency}
\label{sec:exp:efficiency}
We compare the GPU training and test time of \PLACE against all baselines for the three types of ACS queries. 
To facilitate comparison, the training time we report is the elapsed time of one epoch for all learning-based baselines.
Due to space limitation, we report the results for four graphs in Fig.~\ref{fig:efficiency}.

\subsubsection{Training Time}
{Fig.~\ref{fig:afc_training}, ~\subref{fig:afn_training}, and~\subref{fig:ema_training} show the training time of three learning-based approaches.} 
\IACS is the most efficient solution in most cases. \Transzero spends relatively less time on small graphs in \EMA scenarios; however, its training time increases significantly in \Reddit, because its augmented subgraph sampler spends considerably more time to generate community-level subgraphs in large graphs. 
The training time for \AQDGNN is comparable to our \PLACE, though slightly faster. 
Although the computational complexity of \PLACE remains that of the GNN, in practice,  \PLACE still brings additional overhead from constructing query-prompt graphs, leading to a marginally long training time. 
Furthermore, \PLACE can efficiently train large graphs, such as \Reddit, since we sample shards in one epoch instead of training the entire graph, which reflects our scalability. 

\subsubsection{Test time}
Fig.~\ref{fig:afc_test}, ~\subref{fig:afn_test}, and ~\subref{fig:ema_test} compare the average test time of \PLACE with all the baselines. 
In small graphs, the test time of all the approaches is within $10$ milliseconds. In \Reddit, \PLACE needs to make predictions on the entire graph, which results in higher test time. The time of \PLACE for single query on \Reddit remains under one second. \IACS consistently demonstrates the best performance in terms of test time except in \EMA scenario. In \EMA scenario, the algorithmic approach, \CTC requires the least test time in small graphs, as its $k$-truss identification incurs negligible computational cost. 

In summary, despite incurring extra computational overhead in practice, \PLACE outperforms the baselines in prediction accuracy, achieving a favorable balance between computational efficiency and predictive effectiveness.
\begin{table*}[t]
\caption{Performance with Different Types of Prompt Tokens (\%)}
\label{tab:ablation}
\centering
\scriptsize
\resizebox{0.8\textwidth}{!}{
\begin{tabular}{c|c|ccc|ccc|ccc|ccc}
\toprule
\multirow{2}{*}{Setting}             & \multirow{2}{*}{Prompt}       & \multicolumn{3}{c|}{\Cora}                                      & \multicolumn{3}{c|}{\Cornell}                                   & \multicolumn{3}{c|}{\Wisconsin}    & \multicolumn{3}{c}{\Reddit}                              \\
&          & \multicolumn{1}{c}{\Pre} & \multicolumn{1}{c}{\Rec} & \multicolumn{1}{c|}{\Fone} & \multicolumn{1}{c}{\Pre} & \multicolumn{1}{c}{\Rec} & \multicolumn{1}{c|}{\Fone} & \multicolumn{1}{c}{\Pre} & \multicolumn{1}{c}{\Rec} & \multicolumn{1}{c|}{\Fone} & \multicolumn{1}{c}{\Pre} & \multicolumn{1}{c}{\Rec} & \multicolumn{1}{c}{\Fone} \\\midrule
\multirow{4}{*}{\AFC} & w/o attr token 
& 81.46 &94.69 &87.58                 
&80.39 &99.44 &88.91                  
&92.89 &99.19 &95.94        
& 98.64                   & 39.92                      & 56.84                  \\
& w/o virt token 
& 54.11 &98.97 &69.97                     
& 72.43 &99.95 &84.00                  
& 84.40 &99.91 &91.50                  
& 72.34                   & 88.38                      & 79.56                  \\
& w/o both       
& 52.38 &99.35 &68.59                     
& 71.71 &99.95 &83.51                  
& 89.40	&99.94 &94.37               
& 30.89                   & 84.02                      & 45.17                  \\
& full           & 87.34                     & 95.57                      & 91.21                    & 82.46                    & 98.60                       & 90.17                    & 92.56                     & 99.64                      & 95.96                    & 89.87                   & 92.52                      & 91.17                  \\\midrule
\multirow{4}{*}{\AFN} & w/o attr token 
& 87.85 &91.63 &89.70       
&70.38	&96.03 &81.23        
&88.64 &98.33 &93.23        
& 96.05  & 51.99   & 67.46                   \\
& w/o virt token 
& 36.93 &95.63 &53.29       
& 63.67 &99.35 &77.61        
& 85.98 &99.21 &92.12        
& 66.08  & 94.11   & 77.64                   \\
& w/o both       
& 43.57 &97.21 &60.18       
& 58.95 &98.64 &73.80           
& 87.46 &99.36 &93.03        
& 31.70   & 84.46   & 46.10                    \\
& full           
& 85.41       & 96.74        & 90.24       
& 76.40        & 98.38         &85.86       
&89.32 &99.37 &94.06     
& 96.27  & 71.70    & 82.16                   \\\midrule
\multirow{2}{*}{\EMA} & w/o virt token
&57.10 &99.73 &72.62        
&72.11 &99.81 &83.73         
&89.40 &99.94 &94.37     
& 28.90   & 88.34   & 43.55                   \\
& full           & 90.55       & 93.74        & 92.09       & 89.42         & 98.74          & 94.06        & 91.88         & 98.18           & 94.92        & 83.59  & 92.76   & 87.94                  \\\bottomrule   
\end{tabular}}
\end{table*}

\begin{figure}
  \centering
   \includegraphics[width=0.15\textwidth]{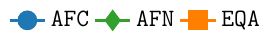}
    \resizebox{0.48\textwidth}{!}{
  \begin{tabular}[h]{c} 
  \hspace{-4ex}
  \subfigure[\Cora]{
    \includegraphics[width=0.162\textwidth]{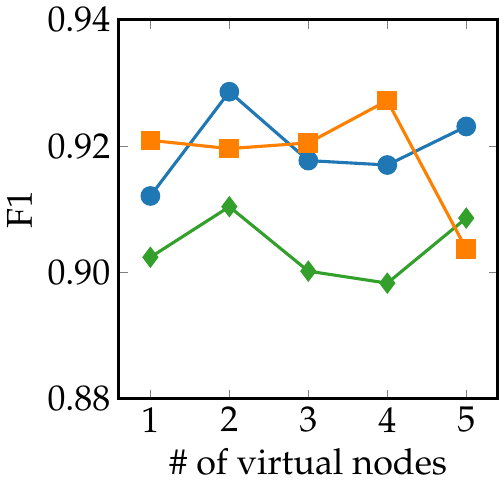}
    \label{fig:vir:cora}}
   \hspace{-2ex} 
  \subfigure[\Cornell]{
    \includegraphics[width=0.15\textwidth]{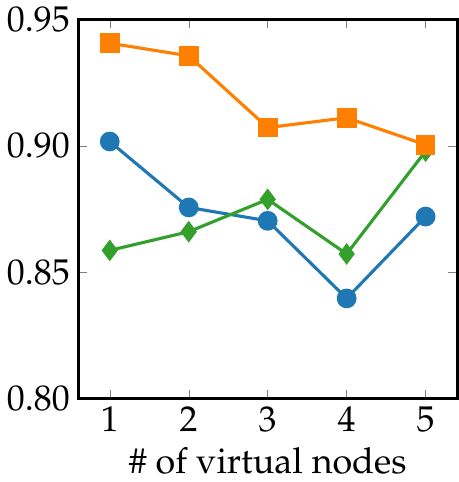}
  \label{fig:vir:cornell}}
  \hspace{-2ex} 
  \subfigure[\Reddit]{
    \includegraphics[width=0.15\textwidth]{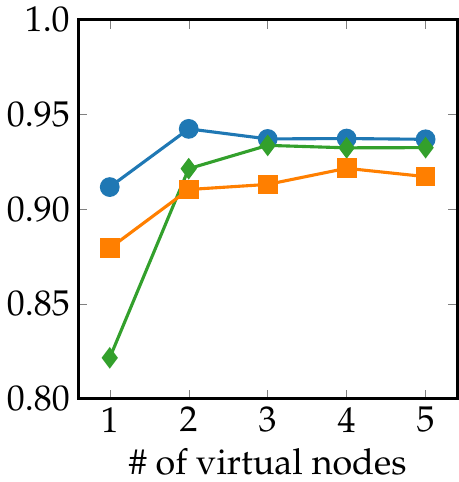}
  \label{fig:vir:reddit}}\\
\hspace{-4ex}
\subfigure[\Cora]{
    \includegraphics[width=0.17\textwidth]{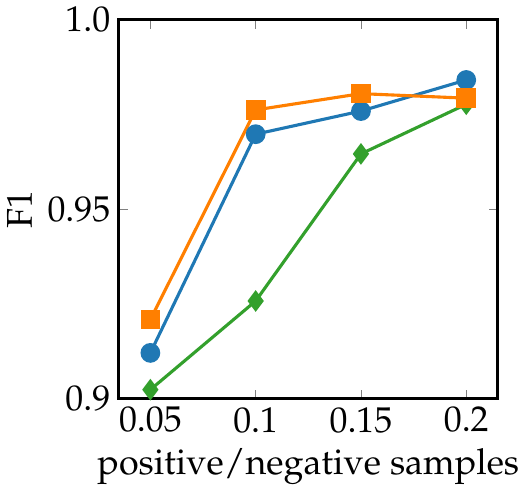}
    \label{fig:posneg:cora}}
   \hspace{-2ex} 
  \subfigure[\Cornell]{
    \includegraphics[width=0.15\textwidth]{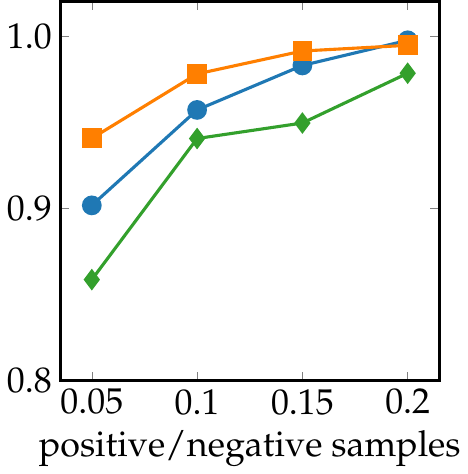}
  \label{fig:posneg:cornell}}
  \hspace{-2ex} 
  \subfigure[\Reddit]{ 
    \includegraphics[width=0.15\textwidth]{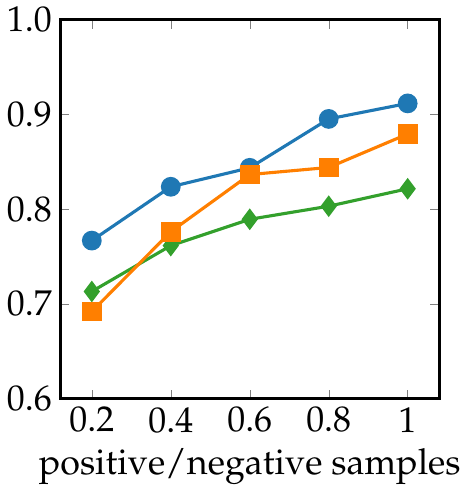} 
  \label{fig:posneg:reddit}}
  \end{tabular}}
   \caption{\Fone Score of \PLACE under Different Number of Virtual Node and Ratio of Positive/Negative Samples}
 \label{fig:virandposneg}
\end{figure}

\subsection{Ablation Study}
\label{sec:exp:ablation}
To verify the effectiveness of both types of prompt tokens, we study different variants of \PLACE w/o prompt tokens.
For \AFC and \AFN, we test three variants: w/o attribute tokens, w/o virtual node tokens,  and w/o both types of prompt tokens, respectively. For \EMA, since queries do not contain query attributes, we test \PLACE w/o virtual node tokens. 
Table~\ref{tab:ablation} presents the performance of these variants.

In general, variants with full prompt tokens consistently achieve the highest \Fone, confirming the  effectiveness of the two types of prompt tokens.  
{We observe that removing virtual node tokens leads to a more significant performance degradation compared to removing attribute tokens except in \Reddit,
indicating that the virtual node prompt has a more substantial impact on the model performance. }
The reason would be that virtual node tokens directly link to query nodes, which promote information propagation near query nodes. 
The alternative training process, which tunes virtual node tokens, also helps the query node update its embedding by incorporating more information relevant to the target community.
Conversely, if the prompt only contains attribute tokens, the graph will lack information pertinent to the query node, preventing the model from identifying the community containing the query node. 
{In large graphs, such as \Reddit, attribute tokens play a more significant role. When training on large graphs, query-route graphs are constructed based on the shards where the query nodes have engaged. However, without attribute tokens, the query attribute information is missing, which negatively impacts performance. This makes attribute tokens more crucial for training on large graphs.}
In \EMA scenarios, removing virtual node tokens leads to significant performance degradation. These tokens are essential to provide query context, and their absence leaves the model regression to a simple baseline of ACS.

\begin{figure}
  \centering
  \includegraphics[width=0.15\textwidth]{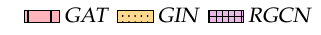}
    \resizebox{0.48\textwidth}{!}{
  \begin{tabular}[h]{c} 
   \hspace{-3ex} 
    \subfigure[\Cora]{
    \includegraphics[width=0.163\textwidth]{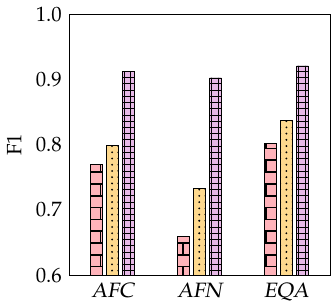}
    \label{fig:gnn:cora}}
   \hspace{-2ex} 
  \subfigure[\Cornell]{
    \includegraphics[width=0.15\textwidth]{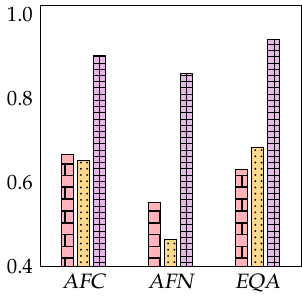}
    \label{fig:gnn:cornell}}
     \hspace{-2ex} 
    \subfigure[\Reddit]{
    \includegraphics[width=0.15\textwidth]{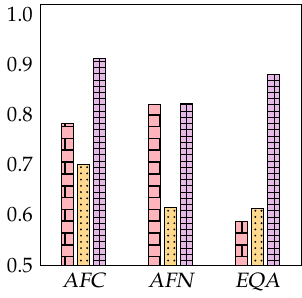}
    \label{fig:gnn:wisconsin}}
\end{tabular}}
   \caption{\Fone Score under Different GNN Layers}
 \label{fig:GNN}
\end{figure}

\begin{figure}
  \centering
  \includegraphics[width=0.4\textwidth]{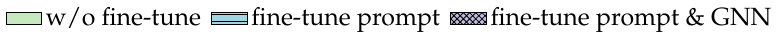}
 \resizebox{0.48\textwidth}{!}{
  \begin{tabular}[h]{c}  
  \hspace{-3ex} 
    \subfigure[Train: \kw{Cor.} Test: \kw{Tex.}]{
    \includegraphics[width=0.165\textwidth]{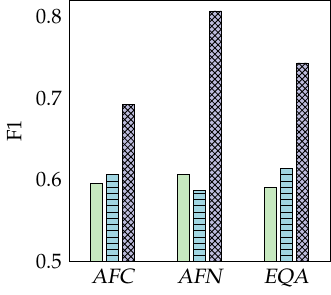}
    \label{fig:transfer:cornelltexas}}
   \hspace{-2ex} 
  \subfigure[Train: \kw{Wis.} Test: \kw{Tex.}]{
    \includegraphics[width=0.15\textwidth]{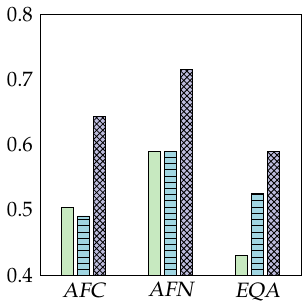}
    \label{fig:transfer:wisconsintexas}}
    \hspace{-2ex} 
    \subfigure[Train: \kw{Cor.} Test: \kw{Was.}]{
    \includegraphics[width=0.155\textwidth]{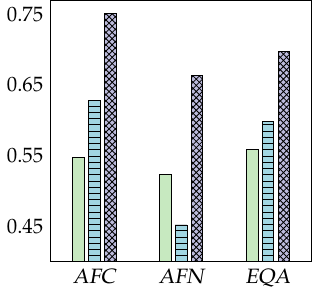}
    \label{fig:transfer:cornellwashington}}
\end{tabular}}
   \caption{Transferability of \PLACE}
 \label{fig:transfer}
\end{figure}

\subsection{Parameter Analysis}
\label{sec:exp:parameteranalysis}

We analyze the parameter sensitivity of \PLACE in three aspects: (1) the number of virtual node tokens,
(2) the ratio of positive/negative samples per training query, and
(3) different GNN layers.

\subsubsection{Different numbers of virtual nodes}
Fig.~\ref{fig:vir:cora}-\subref{fig:vir:reddit} presents the \Fone scores with varying numbers of virtual node tokens, ranging from 1 to 5.
In general, the results indicate that the optimal number of virtual nodes varies across different datasets. For \Reddit dataset, increasing the number of virtual nodes leads to a non-monotonic improvement in \Fone scores across all three scenarios. This improvement is likely because larger graphs need more virtual nodes to capture complex query information.
Considering both efficiency and performance, we set the default number of virtual nodes to $1$ in our experiments. {This choice balances the need for query-specific information with efficiency.}

\subsubsection{Different ratios of positive/negative samples}
We show the \Fone scores with different ratios of positive/negative samples per training query in Fig.~\ref{fig:posneg:cora}-\subref{fig:posneg:reddit}. 
{The training labels vary in the range of $5\% \sim 20\%$, $20\% \sim 100\%$ of the total number of positive/negative samples in small graphs and large graphs, respectively.} %
The \Fone scores show a steady increase as the ratio of positive/negative samples used for training increases.
{Remarkably, our method \PLACE exhibits robust effectiveness in small graphs, maintaining competitive \Fone scores even with as few as $5\%$ of the training samples. 
For large graphs, our method achieves up to $0.7$ \Fone scores with only $20\%$ of the training samples.}
This indicates that \PLACE maintains effectiveness even under circumstances of limited supervision signals.

\subsubsection{Different types of GNN}
We evaluate different GNN layers, specifically \GAT, \GIN, and \RGCN, across three scenarios, as shown in Fig~\ref{fig:GNN}. \RGCN demonstrates the overall best performance across all graphs. 
We introduce edge type information into \RGCN by distinguishing edges in three types: edges in the query-prompt graph, edges in the original graph, and edges across these two graphs. Edges in the query-prompt graph are established based on the pairwise similarity between tokens, which are dynamically changed during training. 
Edges in the original graph are fixed, reflecting its inherent structure, and edges bridging  the two graphs determine how the prompt information influences the original graph. 
\RGCN can better leverage the varying importance of these types of edges to enhance the model performance. \GAT enhances node representation learning through attention mechanisms to weigh node importance. However, it fails to distinguish between prompt tokens and nodes in the original graph, leading to less effective performance compared to \RGCN. 
{The advantage of \GIN lies in the expressivity is as powerful as WL-test, but it also cannot distinguish different types of nodes and edges, resulting in the worse performance compared to \RGCN.} 

\begin{figure}
  \centering
  \includegraphics[width=0.4\textwidth]{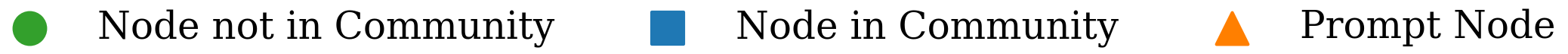}
   \resizebox{0.45\textwidth}{!}{
  \begin{tabular}[h]{c}  
    \subfigure[\Texas: Before]{
    \includegraphics[width=0.2\textwidth]{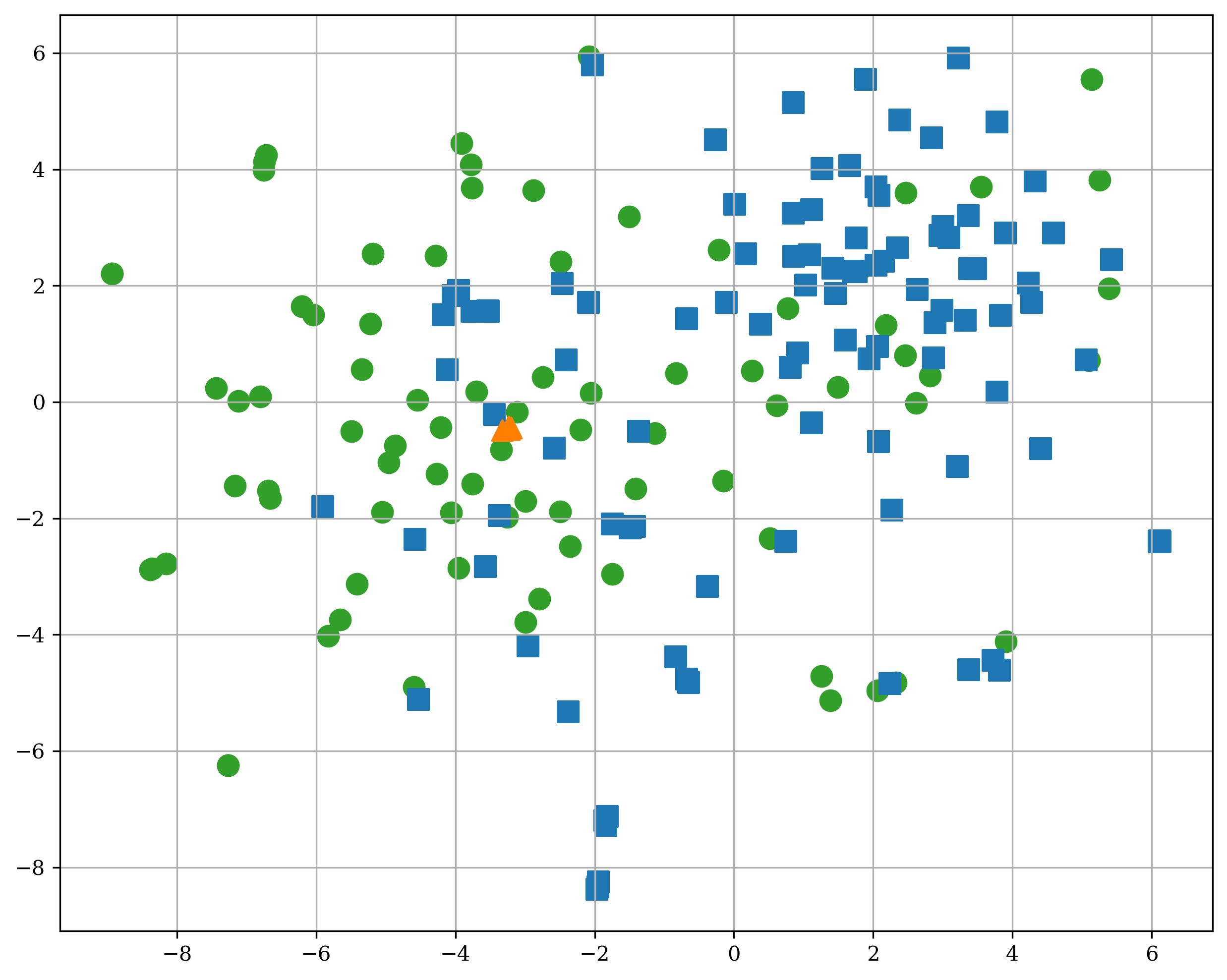}
    \label{fig:case:texasbefore}}
  
  \subfigure[\Texas: After]{
    \includegraphics[width=0.2\textwidth]{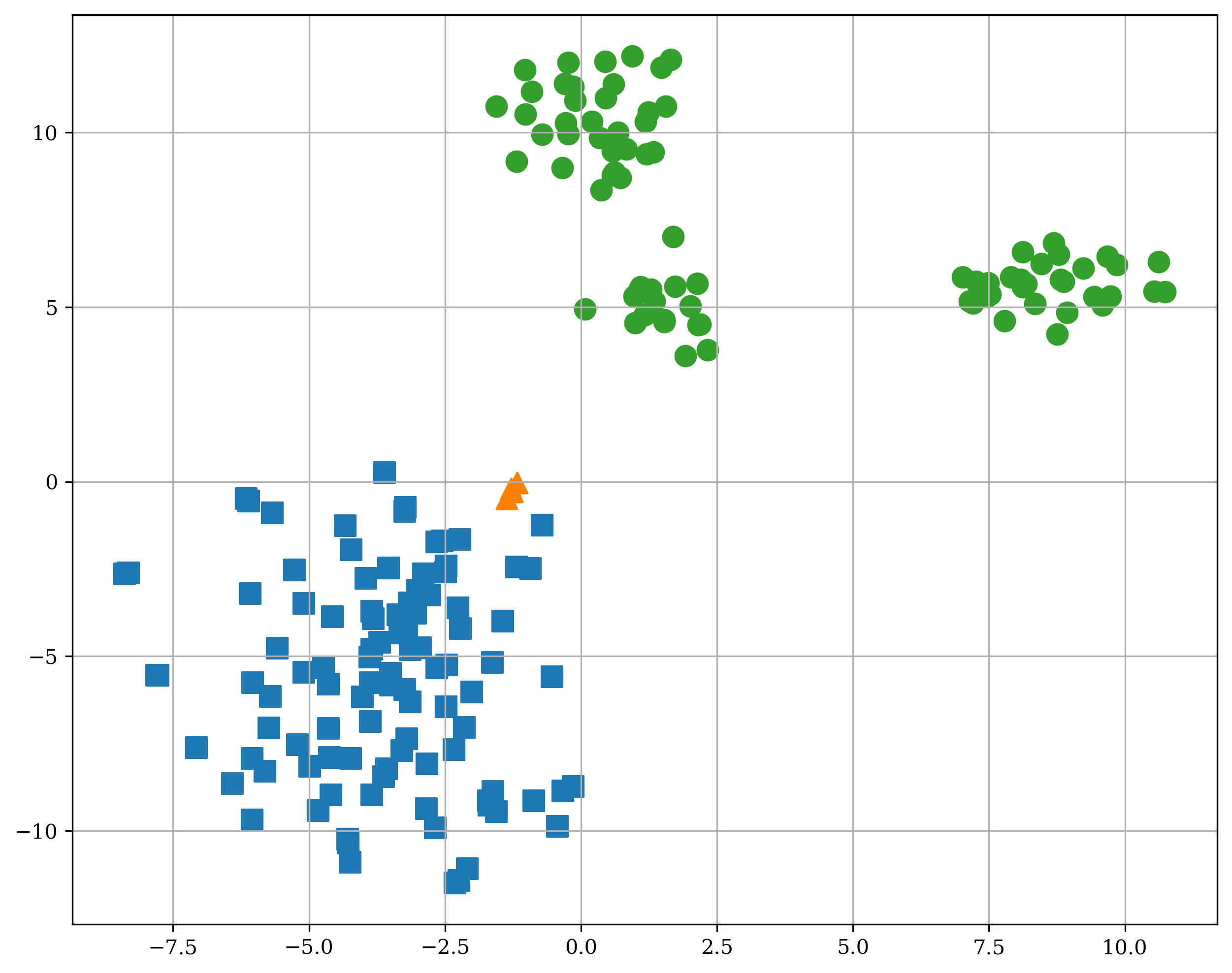}
    \label{fig:case:texasafter}}
     \\
  \subfigure[\Washington: Before]{
    \includegraphics[width=0.2\textwidth]{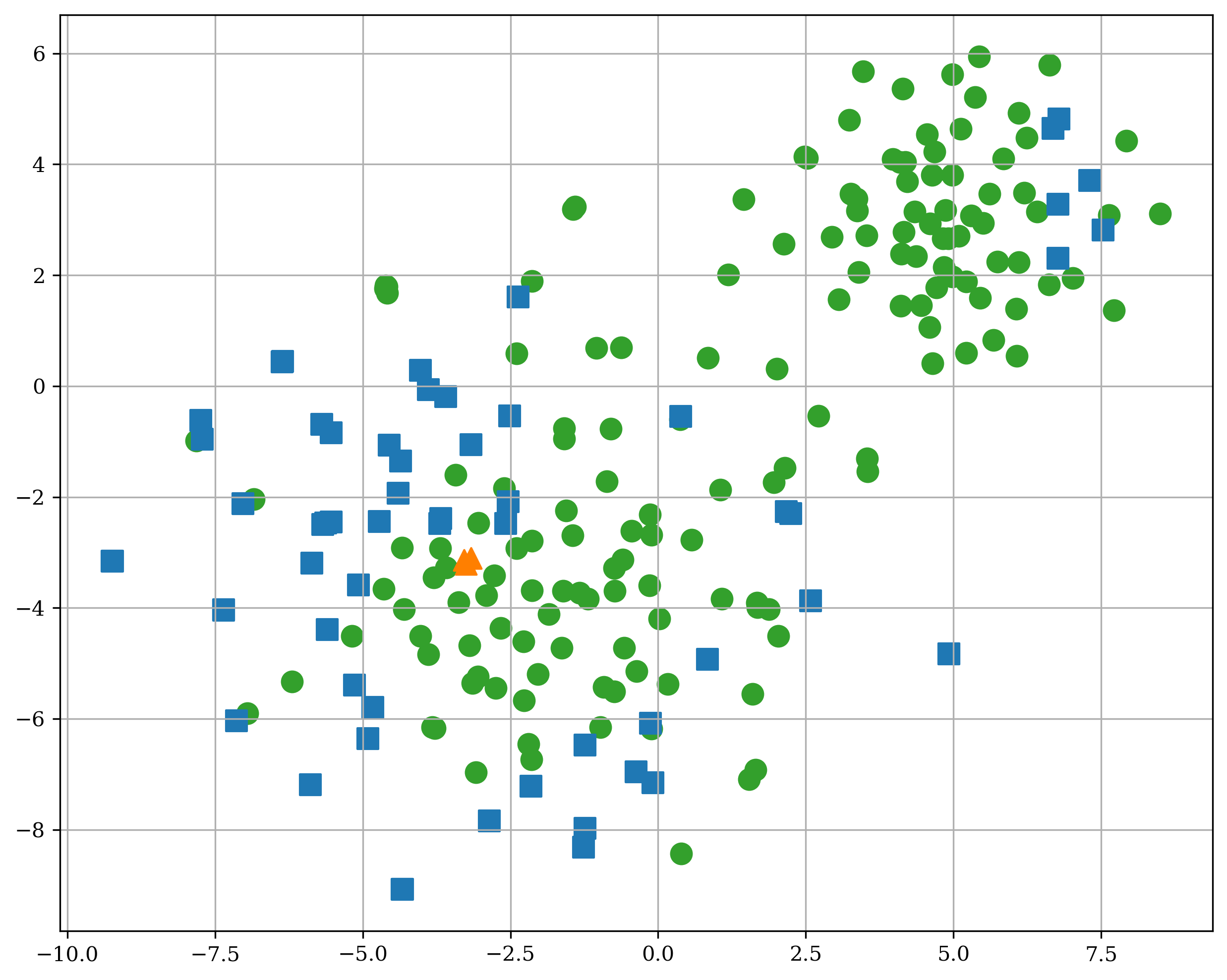}
    \label{fig:case:washingtonbefore}} 
  \subfigure[\Washington: After]{
    \includegraphics[width=0.2\textwidth]{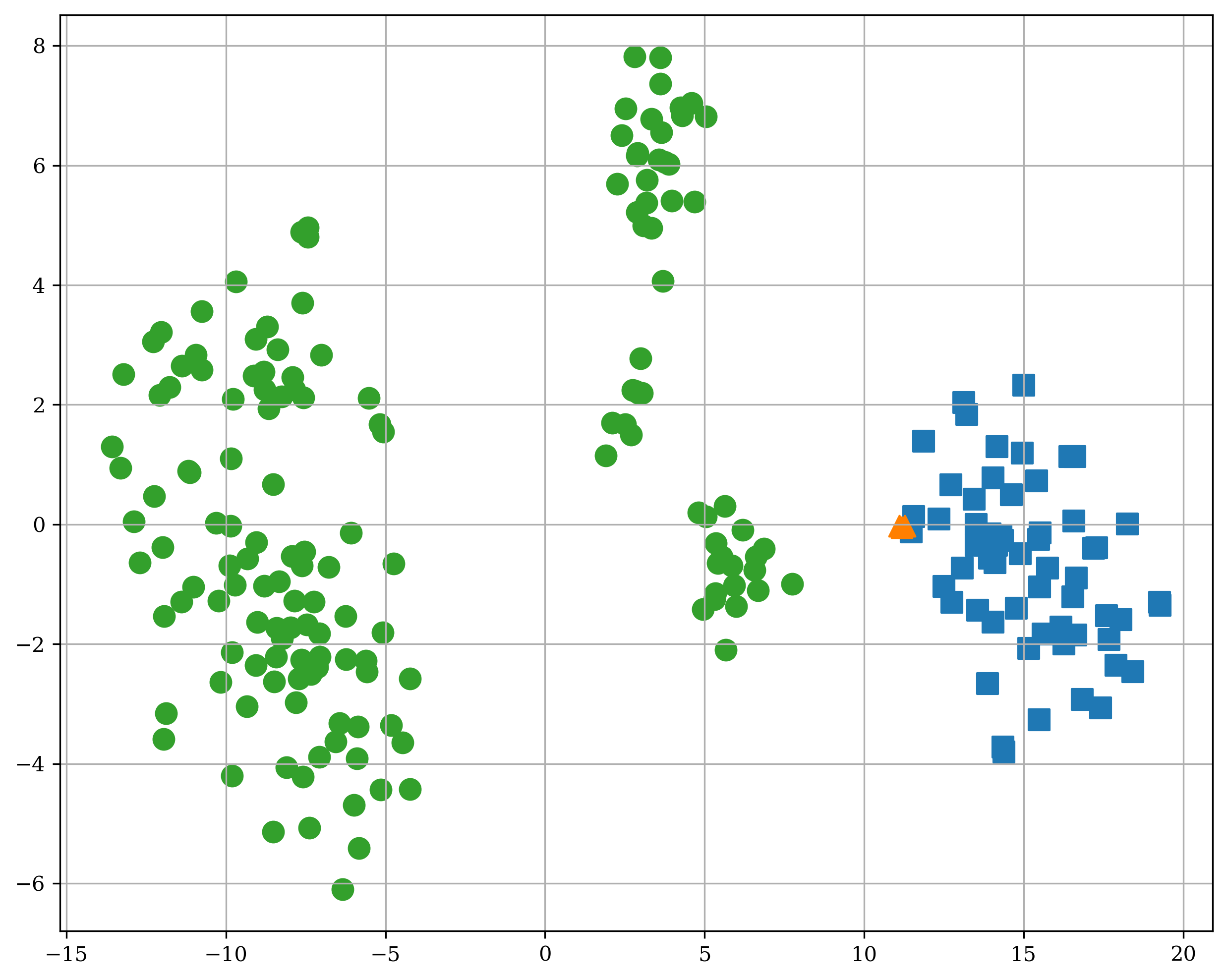}
  \label{fig:case:washingtonafter}}
  \end{tabular}
  }
   \caption{Node Embeddings in prompt-augmented graphs}
 \label{fig:case}
\end{figure}

\subsection{Transferability of \PLACE}
\label{sec:exp:transferability}
The transferability of a GNN model for ACS enables it to generalize to new graphs which differ from the graph used during training. We investigate whether the graph prompt in \PLACE contributes to the transferability.
We train and test \PLACE using different graphs from the WebKB collection, in which the four graphs share the same attribute set and node feature space. 
We report the \Fone under three settings: direct inference (w/o fine-tune), fine-tune prompt only, and fine-tuning both prompt and GNN, as shown in Fig.~\ref{fig:transfer}. 
We set the number of fine-tuning epochs to 200 and randomly draw five queries for fine-tuning in the test datasets.
{The results in Fig.~\ref{fig:transfer} suggest that without fine-tuning, \PLACE still achieves a \Fone of $0.45 \sim 0.6$.} 
In contrast, fine-tuning both the prompt and the GNN exhibits the best \Fone consistently. 
Only fine-tuning the prompt also improves the \Fone to varying degrees in most cases.
These findings demonstrate that the prompt mechanism in \PLACE provides more opportunities for transferring knowledge to different graphs.

\subsection{Case Study}
\label{sec:exp:casestudy}
To explore the functionality of prompt tokens in the ACS task, we conduct a case study visualizing the node embeddings in prompt-augmented graphs. Fig.~\ref{fig:case} presents these node embeddings visualized by t-SNE~\cite{van2008visualizing}. 
Specifically, we visualize node embeddings for the \AFC query in the prompt-augmented graph before training and after training on \Texas (Fig.~\ref{fig:case:texasbefore}-\subref{fig:case:texasafter}) and \Washington (Fig.~\ref{fig:case:washingtonbefore}-\subref{fig:case:washingtonafter}). 
In each figure, green nodes represent those not belonging to the queried community, blue nodes denote those within the queried community, and orange nodes denote prompt tokens.
There are three prompt tokens in each figure: one for the virtual node prompt and two for the query attribute prompt.
Notably, after training, nodes within and without the query community become distinctly separable. Nodes belonging to the same community cluster together and are closer to prompt nodes than other nodes. We speculate that the tuned prompt tokens contain substantial query-related knowledge, enabling the GNN to differentiate nodes effectively by pulling the nodes that belong to the query community closer together while pushing those that do not belong further apart. Additionally, we observe that the distances between the two types of prompt tokens are quite small. 
This may be attributed to the similarity-based edges that facilitate the integration of information among the prompt tokens, forming a cohesive representation of the query context.

\section{Conclusion}
\label{sec:conclusion}
In this paper, we explore the problem of attributed community search using a novel graph prompt learning framework, \PLACE.  
\PLACE overcomes the limitations of existing learning-based methods by dynamically refining the input graph and query through learnable prompt tokens, which are optimized jointly with GNN encoder. A divide-and-conquer strategy is proposed for scalability of \PLACE, enabling the model solve million-scale large graphs. Experimental results in $9$ real-world datasets demonstrate its effectiveness in three query attribute scenarios. \PLACE outperforms the best baseline with higher \Fone scores by $14\%$ on average.


\clearpage


\bibliographystyle{ACM-Reference-Format}
\bibliography{bibfile}

\end{document}